\documentclass[preprint,12pt]{elsarticle}

\usepackage{amssymb}
\usepackage{amsmath}
\usepackage{todonotes}
\usepackage{tcolorbox}
\usepackage{bm}
\usepackage{subfigure}
\usepackage{graphicx}
\usepackage{soul}
\soulregister\cite7
\soulregister\ref7 
\usepackage{multirow}
\usepackage{tikz}
\usepackage{graphicx}
\usetikzlibrary{positioning}
\usepackage{hyperref}
 
\newcommand\red[1]{\textcolor{black}{#1}}

\usepackage[belowskip=2pt,aboveskip=0pt]{caption}

\journal{Knowledge-Based Systems}

\begin{document}

\begin{frontmatter}



\title{Predicting Hate Intensity of Twitter\\ Conversation Threads}


\author[inst1]{Qing Meng\footnote{Both authors contributed equally to this research.}}

\affiliation[inst1]{organization=Hohai University,
            city=Nanjing,
            country=China}

\author[inst2]{Tharun Suresh\footnotemark[1]}
\author[inst4]{Roy Ka-Wei Lee}
\author[inst3]{Tanmoy Chakraborty}

\affiliation[inst2]{organization=IIIT Delhi,
            city=Delhi,
            country=India}

\affiliation[inst3]{organization=Indian Institute of Technology Delhi,
            city=New Delhi,
            country=India}

\affiliation[inst4]{organization=Singapore University of Technology and Design,
            country=Singapore}

\newcommand{\model}{{DRAGNET++}}

\begin{abstract}
Tweets are the most concise form of communication in online social media. Wherein a single tweet has the potential to make or break the discourse of the conversation. Online hate speech is more accessible than ever, and stifling its propagation is of utmost importance for social media companies and users for congenial communication. Most of the research has focused on classifying an individual tweet regardless of the tweet thread/context leading up to that point. One of the classical approaches to curb hate speech is to adopt a \emph{reactive strategy} after the hateful content has been published. This strategy results in neglecting subtle posts that do not show the potential to instigate hate speech on their own but may portend in the subsequent discussion ensuing in the post's replies. In this paper, we propose \model, which aims to predict the intensity of hatred that a tweet can bring in through its reply chain in the future. Our model uses the semantic and propagating structure of the tweet threads to maximize the contextual information leading up to and the fall of hate intensity at each subsequent tweet. We explore three publicly available Twitter datasets -- \emph{Anti-Racism} contains the reply tweets of a collection of social media discourse on racist remarks during US political and COVID-19 background; \emph{Anti-Social} presents a dataset of 40 million tweets amidst the COVID-19 pandemic on anti-social behaviours with custom annotations; and \emph{Anti-Asian} presents Twitter datasets collated based on anti-Asian behaviours during COVID-19 pandemic. All the curated datasets consist of structural graph information of the Tweet threads. We show that \model\ outperforms all the state-of-the-art baselines significantly. It beats the best baseline by an 11\% margin on the Person correlation coefficient and a decrease of 25\% on RMSE for the Anti-Racism dataset with a similar performance on the other two datasets.
\end{abstract}




\end{frontmatter}


\section{Introduction}
\label{sec:introduction}
\newcommand{\model}{{DRAGNET++}}

\textbf{Motivation.} The proliferation of social media has enabled users to share and spread ideas at a prodigious rate. While the information exchanges in social media platforms may improve an individual's social connectedness with online and offline communities, these platforms are increasingly plagued with the rampant onslaught of provocative and toxic content. One such highly toxic content is `hate speech', defined by the Cambridge dictionary as ``\textit{public speech that expresses hate or encourages violence towards a person or group based on race, religion, sex, or sexual orientation }.'' Hate speech on social media has created dissension among online communities and culminated in offline violent hate crimes~\cite{muller2018fanning}. Therefore, addressing the spread of hate speech on social media is critical.

Major social media platforms such as Facebook and Twitter have made significant efforts to combat the spread of hate speech on their platforms~\cite{Times19, bloomberg19}. For example, the platforms have established clear policies regarding hateful conduct~\cite{facebookrules, twitterules}, implemented mechanisms for users to report hate speech, and employed content moderators to detect hate speech. However, such approaches are labor-intensive, time-consuming, and thus not scalable or sustainable in the long run \cite{waseem2016hateful,gamback2017using}. Traditional machine learning and deep learning methods have also been proposed to automatically detect hate speech in online social media~\cite{schmidt_survey_2017,fortuna2018survey,poletto_resources_2021}. However, most of the existing methods are limited to classifying hate speech at the individual post level, ignoring the network and propagation effects of hateful content on social media~\cite{liu_forecasting_2018}. 

\begin{figure}[t]
    \centering
	\includegraphics[scale=0.6]{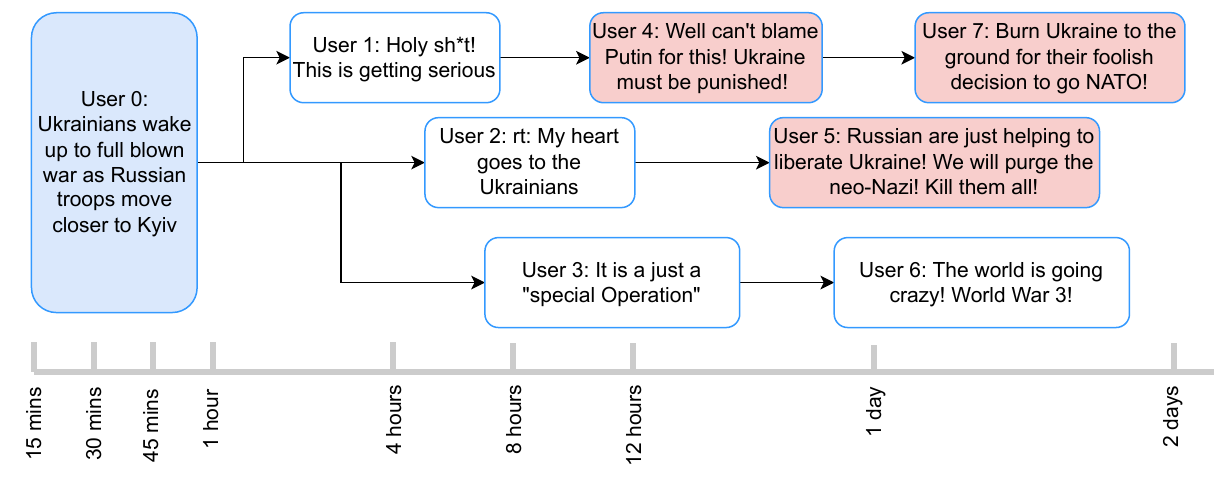}
\caption{An example of a tweet propagation. The hateful tweets are highlighted in red.}
\label{fig:prop_example}
\vspace{3mm}
\end{figure}

Ideally, social media content moderators would want to identify hateful posts and monitor posts and threads more likely to incite hatred. Consider the example of tweet propagation shown in Figure~\ref{fig:prop_example}. The initial source tweet is a benign tweet that reports the Russian invasion of Ukraine. However, as time evolves, we observe hateful content justifying the Russian invasion and promoting violence against the Ukrainians. Existing automated hate speech detection methods could help content moderators detect hateful tweets. Nonetheless, content moderators could significantly improve the effectiveness and efficiency of moderation by prioritizing posts likely to generate a large number of hateful tweets in their retweet or reply threads. This a challenging problem as the post that induces hateful content may not be hateful itself (e.g., source tweet in Figure~\ref{fig:prop_example}). Early detection of tweets likely to generate a large amount of hateful content becomes a critical real-world issue. 

Very few studies examined the hate intensity of Twitter conversation threads for early hate speech detection.  \citet{lin2021early} manually categorized Twitter conversation threads into different levels of hatred and proposed a deep learning model to forecast and classify tweets into the pre-defined hatred levels. Recently, \citet{sahnan2021better} introduced the hate intensity prediction task and proposed DRAGNET, a deep stratified learning framework that predicted the intensity of hatred that a root tweet can fetch through its subsequent replies. Nevertheless, DRAGNET assumes a linear sequence for the conversation thread in its model, in which tweets in conversation threads are arranged in a chronological sequence. This neglects the tree structural information inherently present in a conversation thread; there could be multiple branches in a conversation thread similar to the example shown in Fig.~\ref{fig:prop_example}. The structural information may be an important feature that could better inform and improve hate intensity prediction. For instance, we may notice a heated debate along the branches of the conversation threads; therefore, modeling the dynamics of the tree structure of the conversation thread may aid us in better predicting the eventual hate intensity of the conversation thread.

\textbf{Research Objectives.}    In this paper, we aim to fill this research gap and extend our earlier work \cite{sahnan2021better} by proposing \model\ that models multifaceted information to forecast the hate intensity of a Twitter conversation thread. Specifically, we define our problem statement as follows -- \red{\emph{given a root tweet and a few of its initial replies, can we predict the hate intensity of the subsequent replies of the tweet}? To model this problem, we first quantify the hate intensities of tweets in the conversation thread and express them as a series of hate intensities, transforming the task into a time-series problem.}  At a high level, \model\ adopts a similar stratified learning framework as suggested in \cite{sahnan2021better} to model the hate intensity profiles of Twitter conversation threads and categorize them into clusters of varying hate intensity. In addition to the content sentiment and temporal information, \model\ also captures structural information of the conversation thread. Specifically, \model\ adopts  Graph Neural Networks (GNN) to learn the semantic and propagation structure of conversation threads.

As the prediction of hate intensity in conversation threads is a relatively new research problem, we conduct thorough experiments and analyses to evaluate \model. We collate three publicly available real-world datasets -- \emph{Anti-Racism} with ~$3.5k$ root tweets, \emph{Anti-Social} with ~$668k$ root tweets, and \emph{Anti-Asian} with $218k$ root tweets. We extensively analyze these datasets to examine the hate intensities of real-world Twitter conversation threads. We benchmark \model\ against DRAGNET and six other baselines on the hate intensity task. Finally, we examine case studies and conduct ablation studies to explain the advantages and limitations of \model.

\textbf{Contributions.} We summarize our contributions below:
\begin{itemize}
    \item We analyze the hate intensity in three large-scale real-world Twitter datasets. This is the first large-scale study that examines hate intensity in Twitter conversation threads.
    \item We propose \model, a stratified learning framework with structural graph augmentation to perform hate intensity prediction.
    \item We conduct extensive experiments and show that \model\ consistently outperforms state-of-the-art methods in the early prediction of hateful conversation threads. Specifically, \model\ outperforms the best baseline by at least $34\%$ on  Pearson Correlation Coefficient (PCC), reduction of at least $200\%$ on Root Mean Square Error (RMSE), and $7\%$ lower on Mean Forecast Error (MFE) across the datasets. Comparing the performance of \model\ with DRAGNET shows an average improvement of $8.33\%$ on PCC and an average reduction of $16.07\%$, and $18.20\%$ on RMSE and MFE, respectively.
    \item Our case studies demonstrate the viability of predicting hate speech early to prevent the propagation of hateful content.
\end{itemize}

{\bf Organization of the paper:} We start by presenting the developments until recently related to hate speech detection and time series forecasting in Section~\ref{sec:related}. 
Following this, we define the formulation of hate intensity profiles appropriate to our problem statement in Section~\ref{sec:preliminaries}. We then move on to detailing our model in Section~\ref{sec:model}. Further, we analyze the various datasets and present their statistics and observations in Section ~\ref{sec:dataset}. We discuss a brief overview of various baselines in Section~\ref{sec:baselines}. Section ~\ref{sec:experimental_analysis} presents a detailed comparison, analysis of the model performance, and the effect of various parameters used to fine-tune the model. We conclude the paper in Section \ref{sec:conclusion}.

\textbf{Reproducibility}. The source codes of \model\ and all the baselines and the dataset are available at the following link: \url{https://github.com/LCS2-IIITD/Predicting-Hate-Intensity}.

\section{Related Work}
\label{sec:related}

\subsection{Causes of Hate Speech}

\red{There are several reasons for individuals to harbor hatred towards each other or a particular group of people. According to Navarro's \cite{Navarro-2013-Hatred} discussion, hate can arise when someone perpetrates harm or discrimination against others. Hate is often characterized by the devaluation of the victim, which can escalate to the point of elimination in extreme cases. This notion demonstrates the snowball effect commonly associated with hatred as arguments progress. In the context of the internet, John et al. \cite{Suler_Cyber_2004} highlight that people tend to self-disclose more frequently or intensely in online media, a phenomenon known as the online disinhibition effect. Furthermore, Lindsay et al. \cite{Lindsay_Maizland_2019_cfr} argue that social media, a highly interconnected platform, is driven by algorithms and primarily profits from user engagement. As a result, it unintentionally propagates extreme content under the guise of serving user interests. The causes of hate speech are multifaceted, and some of them are introduced due to technological advancements. While we do not explicitly study the cause of hate speech in individual tweets, our proposed model possesses the capability to identify "high-risk" tweets and reply chains, alerting content moderators to conduct further analysis and gain a deeper understanding of the causes of hate speech.} 

\subsection{Hate Speech Detection in Social Media}

Industry and academics have paid close attention to the work on hate speech detection. With the advancement of deep learning, neural language processing systems that can automatically extract features from the text have seen a lot of success~\cite{kim2014convolutional, sutskever2014sequence, vaswani2017attention}. This opens up new possibilities for hate speech detection~\cite{mehdad2016characters, badjatiya2017deep, zhang2018detecting, arango2019hate, founta2019unified,awal2023model}. Gamback et al.~\cite{gamback2017using}, for example, used Convolutional Neural Networks (CNN) to classify hate speech by extracting word similarities. Similar research was done by Park et al.~\cite{park2017one}, who presented the HybridCNN model to investigate word and character combination patterns to detect hate speech. On the other hand, Del et al.~\cite{del2017hate} used Long Short Term Memory (LSTM) to record the long-term dependencies of words in phrases to differentiate hatred remarks. Badjatiya et al.~\cite{badjatiya2017deep} evaluated the use of the LSTM model in conjunction with the Gradient-Boost Decision Tree (GBDT) to perform hate speech classification and found that it greatly enhanced performance. Zhang et al.~\cite{zhang2018detecting} presented a CNN+GRU network architecture to investigate word dependency for recognizing hate speech tweets, combining the benefits of CNN-based models and Gate Recurrent Unit (GRU)-based models. However, most research focuses on learning a particular textual property while ignoring other valuable data. Cao et al.~\cite{cao2020deephate} introduced DeepHate, a deep learning-based hate speech detection algorithm for mining multi-faceted textual representations. Lee et al.~\cite{lee2021disentangling} later developed DisMultiHate, a hateful meme classification model that can learn both textual and visual information. Recently, Awal et al. \cite{awal2023model} proposed a meta-learning-based framework (HateMAML) that can effectively detect hate speech in eight different low-resource languages.\par

Pre-trained language models such as BERT~\cite{devlin2018bert} and GPT \cite{radford2018improving} can extract external information from massive volumes of text data. These models have also been used in hate speech detection models with promising results \cite{mozafari2019bert, awal2021angrybert}. For example, Awal et al.~\cite{awal2021angrybert} used the pre-trained BERT model as a shared layer and created AngryBERT, a multi-task learning model that can jointly detect hate speech and classify sentiment. Nonetheless, most previous research works have focused on categorizing hate speech. A few research looked into how hate speech spreads through social media. In a recent discussion, Dahiya et al.~\cite{dahiya_would_2021} discussed the necessity of anticipating the hatred intensity of tweets. Lin et al.~\cite{lin2021early} classified the tweets into different levels of hatred and suggested that the HEAR model tracks posts likely to cause hate speech. Closer to our work is DRAGNET~\cite{sahnan2021better}, which is a deep stratified learning framework that predicts the hate intensity of a conversation thread based on what a root tweet can fetch through its subsequent replies. DRAGNET models the linear sequence for the conversation thread chain, where the tweets are arranged chronologically. Such a modeling approach ignores the conversation thread's inherent structural information propagation information. We address this limitation and propose \model which considers the Twitter conversation thread's structural information to improve hate intensity prediction.

On the other hand, some comprehensive datasets are proposed in hate speech shared tasks. They focus on niche aspects of hate speech. For example, \citet{basile-etal-2019-semeval} released a task focusing on hate speech against immigrants and women in a multilingual setup. \citet{sanguinetti-etal-2020} formulated testing out of domain dataset between tweets and news headlines. They explored the prevalence of verbless fragments in most hate speech texts. Furthermore, potential hate speech spreaders can be inferred from an individual's Twitter feed \citep{Rangel2021ProfilingHS}. Narrowing a step deeper, \citet{pavlopoulos-etal-2021-semeval} discussed toxic span detection from previously annotated hateful comments by re-annotating them at the span level. They gave a more fine-grained understanding of spans contributing to toxicity.

\subsection{Time Series Forecasting Modeling}
The hatred intensity prediction problem can be reduced to a time series prediction task, where we forecast the hate intensity of the conversation thread in the future. Time series models are designed to predict values or trends over time in the future and have been extensively studied in many fields such as transportation\cite{ma2020forecasting}, finance\cite{andersen2005volatility}, event forecasting, \cite{ding2019modeling} and disease transmission \cite{matsubara2014funnel}. Therefore, we also introduce the existing studies of time series forecasting (TSF) models.

Traditional TSF methods, such as ARMA \cite{rojas2008soft}, exponential smoothing \cite{gardner1985exponential}, and linear space models, have been the basis of much work and achieved better performance. With the rapid development of deep learning technology in recent years, more and more end-to-end TSF models have been proposed. Compared to traditional methods, deep learning-based models such as CNN, RNN, and LSTM can automatically extract features from the input without domain expertise, widely accepted in many fields. For example, Oord et al. \cite{oord2016wavenet} adopted CNN in raw audio generation and proposed the WaveNet model. Based on WaveNet, Borovykh et al. \cite{borovykh2017conditional} utilized CNN to perform conditional time series forecasting tasks. The model shared filter weights assuming the hidden patterns are time-invariant at each step. In addition, the RNN contains an internal memory state, which can also retain information about previous time steps \cite{wei2021machine}. However, due to vanishing gradients, the performance of classical RNN-based models degrades as the sequence length increases. Later, the LSTM model overcomes the long-term dependency problem to some extent. Elsworth et al. \cite{elsworth2020time} used LSTM for TSF tasks, improving performance and robustness. 

However, most of the above models only focus on one-step predictions. Some recent studies \cite{mariet2019foundations, fan2019multi} generalize sequence-to-sequence (Seq2Seq) models and propose them for multi-step time series forecasting methods. Fan et al. \cite{fan2019multi} exploited attention mechanisms to capture the temporal context information extracted by the RNN encoder. Combined with a bidirectional LSTM decoder, the model can generate multiple future horizons simultaneously. Later, researchers proposed new architectures \cite{sen2019think, li2019enhancing} to reduce error accumulation in multi-step prediction. For instance, Sen et al. \cite{sen2019think} combined global matrix factorization models and local temporal networks to capture latent patterns in time series. In recent years, the Transformer architecture has been proposed for natural language processing tasks and has also achieved success on time series data \cite{li2019enhancing, oreshkin2019n}. Moreover, some studies \cite{yuan2019diverse, salinas2020deepar} proposed models to estimate the probability distribution of future time series. For example, Yuan et al. \cite{yuan2019diverse} and Koonchli et al. \cite{koochali2019probabilistic} employed Generative Adversarial Networks (GANs) to predict future values. Salinas et al. \cite{salinas2020deepar} proposed a DeepAR model for probabilistic prediction based on auto-regressive recurrent networks.

\section{Preliminaries}
\label{sec:preliminaries}
\begin{table}[!th]\small
    \caption{Important notations and denotations.}
    \begin{tabular}{p{0.15\textwidth}p{0.85\textwidth}}
        \hline
         {\bf Symbol} & {\bf Definition}  \\
         \hline
         $\varphi$ & Root tweet\\
          $c_i^{\varphi}$ & $i^{th}$ reply to root tweet $\varphi$\\
          $R_{p,q}$ & Hate value set of $q$ reply tweets of the $p^{th}$ root tweet\\
          $R^*_{p,q}$ & The set of hate values for the $q$ reply tweets of the $p^{th}$ root tweet predicted by the autoencoder\\
          $S_{s_{(1,n)}}$ & Cosine similarity set between root tweet and $n$ reply tweets \\
          $R_{s_{(1,n)}}$ & Set of hate intensity profiles\\
          $j$ & Number of clusters\\
          $\delta$ & Window size\\
          $t_h$ & Number of replies in history\\
          $t_f$ & Index of the last reply tweet in the conversation thread\\
          $n$ & Maximum length of the conversation thread\\
          $s$ & Total number of conversation threads\\
          $N_{X_h}$ & Dimension of encoded history latent vector\\
          $N_{X_f}$ & Dimension of encoded future latent vector\\
        $\mathcal{X}$ & Latent vector of $R_{p,q}$\\
        $\mathcal{X}^*$ & Latent vector of $R^*_{p,q}$\\
        $C_c$ & List of cluster centers\\
        $\mathcal{P}(C_{ci})$ & Likelihood of belongingness to $i^{th}$ cluster identified with $C_{ci}$\\
        $\mathcal{P}^*(C_{ci})$ & Predicted weight for $i^{th}$ cluster centre $C_{ci}$ to calculate $\mathcal{X}_c$\\
        $\mathcal{X}_d$ & Pre-processed prior vector\\
         $\mathcal{X}_{h}$, $\mathcal{X}_{f}$ & The representation of historical and future conversation threads \\
         $\mathcal{X}_{h}^c$, $\mathcal{X}_{f}^c$ & The representation of pseudo historical and pseudo future conversation threads \\
         $\mathcal{E}_h(\cdot)$ & The encoder model of historical conversation threads \\
         $\mathcal{E}_f(\cdot)$ & The encoder model of future conversation threads\\
         $\mathcal{GM}(\cdot)$ & \red{Fuzzy Clustering}\\
         $\mathcal{PR}(\cdot)$ & \red{Prior model}\\
         $\mathcal{FP}(\cdot)$ & \red{Future Predictor}\\
         $\mathcal{FP}_d(\cdot)$ & $1^{st}$ segment of \red{Future Predictor}\\
         $\mathcal{FP}_p(\cdot)$ & $2^{nd}$ \red{Future Predictor}\\
         $\mathcal{D}(\cdot)$ & \red{Decoder} \\
          $\mathcal{TE}(\cdot)$ & \red{Tree Encoder}\\ 
         \hline
    \end{tabular}
    \label{tab:Notations}
\end{table}




\textbf{Hate Intensity Definition.} Table \ref{tab:Notations} summarizes the denotations of the important notations. Let an ordered sequence of first $t$ number of replies to a root tweet $\varphi$ be $\mathcal{T}_{1,t}^{\varphi}=<c_1^{\varphi}, c_2^{\varphi}, \cdots , c_t^{\varphi}>$, where $c_i^{\varphi}$ refers to the $i^{th}$ reply. As the replies in the original dataset have no ground-truth hate intensity, we quantify the hate intensity of each reply using a weighted sum of two measures as suggested in \cite{dahiya_would_2021}:
\begin{equation}
    \mathcal{H} = w \mathcal{H}_c(c)+(1-w) \mathcal{H}_l(c),
\end{equation}
where $\mathcal{H}_c$ indicates the probability that the reply is hateful, and it is calculated by a state-of-the-art hate speech detection model (we will elaborate it in Section \ref{sec:experimental_analysis}). $\mathcal{H}_l$ is the average score for all words in a reply from a model-independent hate lexicon that comprises 2,895 words as proposed in \cite{wiegand2019inducing}. $w \in [0,1]$ is a hyper-parameter that balances two quantities of the hate intensity score. \red{As $\mathcal{H}_c \in [0,1]$ and $\mathcal{H}_l \in [0,1]$, the final hate intensity score of a reply $\mathcal{H}$ is still between 0 and 1. We do not filter any tweet threads based on their hate intensity}. Therefore, each conversation thread $\mathcal{T}$ can be mapped to a sequence of hate intensity scores, 
\begin{equation}
    \mathcal{H}(\mathcal{T}_{1,t}^{\varphi}) = <\mathcal{H}(c_1^{\varphi}), \mathcal{H}(c_2^{\varphi}), \cdots, \mathcal{H}(c_t^{\varphi})>
\end{equation}

To avoid noise and drastic fluctuations that are not in line with reality,  we further smooth the hate intensity sequence by utilizing a \textit{rolling average operation} with window size $\delta$ \cite{dahiya_would_2021}. A window consists of $\delta$ consecutive replies, and the window that starts from the $k^{th}$ reply is denoted by $\mathcal{T}_{k,k+\delta}^{\varphi}$. Finally, the hate intensity of a window for the tweet $\varphi$ is measured as,
\begin{equation}
\begin{aligned}
    \mathcal{H}(\mathcal{T}_{k,k+\delta}^{\varphi}) &= \sum_{c \in \mathcal{T}_{k,k+\delta}^{\varphi}}\mathcal{H}(c)
     = w \sum_{c \in \mathcal{T}_{k,k+\delta}^{\varphi}} \mathcal{H}_c(c)+(1-w) \sum_{c \in \mathcal{T}_{k,k+\delta}^{\varphi}}\mathcal{H}_l(c),
\end{aligned}
\label{equ:hate_intensity}
\end{equation}
where $\mathcal{H}(\mathcal{T}_{k,k+\delta}^{\varphi}) \in [0, \delta]$, and $\delta$ is the window size. 



\textbf{Sentiment Features.} The emotional feedback of users is reflected in the sentiment features of reply posts, which provide both for and against arguments of the original tweet. Therefore, we use the cosine similarity between the sentiment embedding of the root tweet ($\varphi$) and its accompanying replies ($c^{\varphi}_1, c^{\varphi}_2, \cdots$) to capture the sentiment context of a conversation thread, denoted as $\mathcal{CS}(c_i) = CosineSim(Embed(c^\varphi_i), Embed(\varphi))$. The sentiment embedding is the second last fully-connected layer from the pre-trained \textit{XLNet} model \cite{yang2019xlnet} for sentiment classification. To smooth the value and eliminate the effect of noise, we also apply the \textit{rolling average operation} to the sentiment context sequences $\mathcal{CS}(\mathcal{T}^{\varphi}_{1,t})$ with the same window size $\delta$ as performed on $\mathcal{H}(\mathcal{T}_{1,t}^{\varphi})$.

\textbf{Problem Definition.} Given (i) a root retweet $\varphi$, (ii) its last $t_h$ historical replies $\mathcal{T}_{1,t_h}^{\varphi}=\{c_1^{\varphi},c_2^{\varphi},\cdots,c_{t_h}^{\varphi}\}$, (iii) the corresponding hate intensity sequence $\{\mathcal{H}(\mathcal{T}_{k,k+\delta}^{\varphi})|k \in [1,2,\cdots,t_h-\delta]\}$, and (iv) $G_{\varphi} = (V, E)$  referring to the adjacency matrix of the propagation tree, where $V=\{\varphi, c_1^{\varphi},c_2^{\varphi},\cdots,c_t^{\varphi}\}$, and $e_{ij} \in \{0,1\}$ denote the retweeting/reply relationships between tweets (i.e., root tweet or replies), we aim to predict the hate intensity of the upcoming replies $c_{t'}^{\varphi}$ in the propagation tree of the root tweet $\varphi$. However, in corresponding to the historical hate intensity sequence, we consider predicting the hate intensity for each window of $c_{t'}^{\varphi}$, denoted by $\mathcal{H}(\mathcal{T}_{t',t'+\delta}^{\varphi})$, instead of directly predicting the hate intensity of each reply.

\section{Methodology}
\label{sec:model}

This section presents our proposed hate intensity prediction method, {\model}. It is a deep stratified learning \cite{hastings2016stratified} method that splits heterogeneous data points (in this case, Twitter conversations) into homogeneous clusters/strata before training a deep regressor on each stratum to predict hate intensity. 

\begin{figure}[!t]
    \centering
    \includegraphics[width=\textwidth]{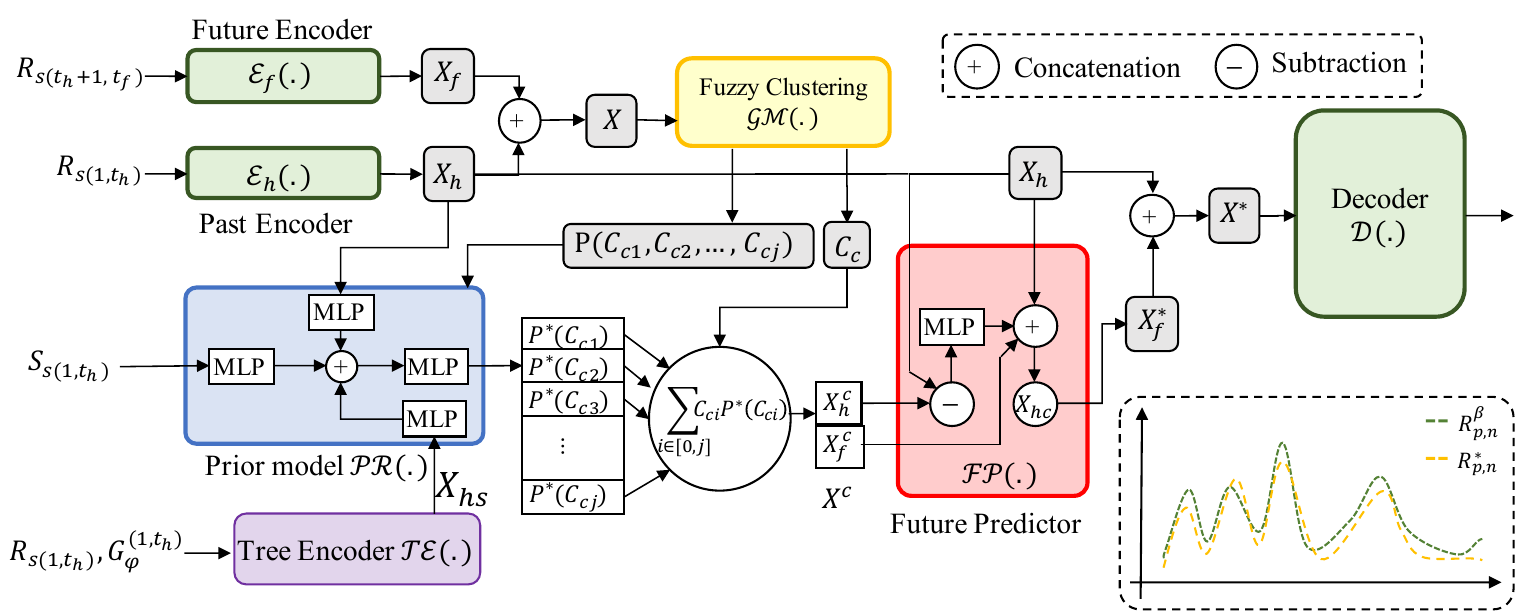}
    \vspace{-5mm}
    \caption{The overall framework of \model. The autoencoder is trained on hate-intensity profiles of the entire conversation thread. Using the trained autoencoder, the history and future latent representations are concatenated and clustered using the fuzzy clustering algorithm $\mathcal{GM(\cdot)}$. In the Prior model, $\mathcal{PR(\cdot)}$, graph representations, history latent representations and sentiment features are concatenated to generate the Prior Knowledge vector. On inference, (a) the history latent representation, (b) sentiment similarity features of the history, and (c) the graph representation of the conversation thread are used to predict the future hate intensity profile.}
    \label{fig:framework}
    \vspace{5mm}
\end{figure}

Figure \ref{fig:framework} illustrates the overall architecture of {\model}. \red{The two-dimensional vector forms the training set of window-wise hate intensity profile and sentiment context value sequences. }

\begin{equation}
    \begin{aligned}
    \mathcal{R}_{s_{(1,n)}} &= \{\mathcal{R}_{(p,q)} : 1 \leq p \leq s, 1 \leq q \leq n\}\\
    \mathcal{R}_{(p,q)} &= \{\mathcal{H}(\mathcal{T}^{\varphi_p}_{k, k+\delta}) : 1 \leq k \leq q - \delta \}\\
    \mathcal{S}_{s_{(1,n)}} &= \{\mathcal{S}_{(p,q)} : 1 \leq p \leq s, 1 \leq q \leq n\}\\
    \mathcal{S}_{(p,q)} &= \{\mathcal{CS}(\mathcal{T}^{\varphi_p}_{k, k+\delta}) : 1 \leq k \leq q - \delta \}\\
    \end{aligned}
    \label{equ:hate_intensity_profile}
\end{equation}
\red{where $s$ represents the total number of conversations, and $n$ represents the maximum conversation length. The elements $\mathcal{R}_{(p,q)} \in \mathcal{R}_{s_{(1,n)}}$ and $\mathcal{S}_{(p,q)} \in \mathcal{S}_{s_{(1,n)}}$ represent the $p^{th}$ data point, where the conversation thread is of length $q$ ($\varphi_p$ represents the $p^{th}$ root tweet).}

{\model} uses an autoencoder to learn low-dimensional latent representations for the hate intensity profile of conversation threads. Specifically, the model learns two alternative latent representations: $\mathcal{X}_h$, which represents the first few replies, and $\mathcal{X}_f$, which represents the future hatred trend for the remaining replies. Once these representations, $\mathcal{X}_h$ and $\mathcal{X}_{f}$, are combined, the model uses an unsupervised setting to apply a fuzzy clustering technique to give cluster membership probabilities and cluster centers to each conversation thread. The number of clusters is determined by a hyper-parameter $j$. 
Following this, the model  trains with historical node and propagation structure $G_{\varphi}^{(1,t_h)}$ using the tree encoder to learn the structure embedding $\mathcal{X}_{hs}$. The model then trains a new deep neural network unit to predict cluster membership probabilities given $\mathcal{X}_h$, $\mathcal{X}_{hs}$ and $\mathcal{S}_{s_{(1,t_h)}}$, which assign cluster centers for a new conversation thread.
Finally, using $\mathcal{X}_h$ and $\mathcal{P}(C_{c1}, C_{c2}, \cdots, C_{cj})$, a novel deep regressor predicts the latent representation of the future hate trend $\mathcal{X}^*_f$, which, when coupled with $\mathcal{X}_h$, is transformed to the whole hate trend by the decoder trained during the autoencoder phase.

\begin{figure}[!t]
    \centering
    \includegraphics[width=\textwidth]{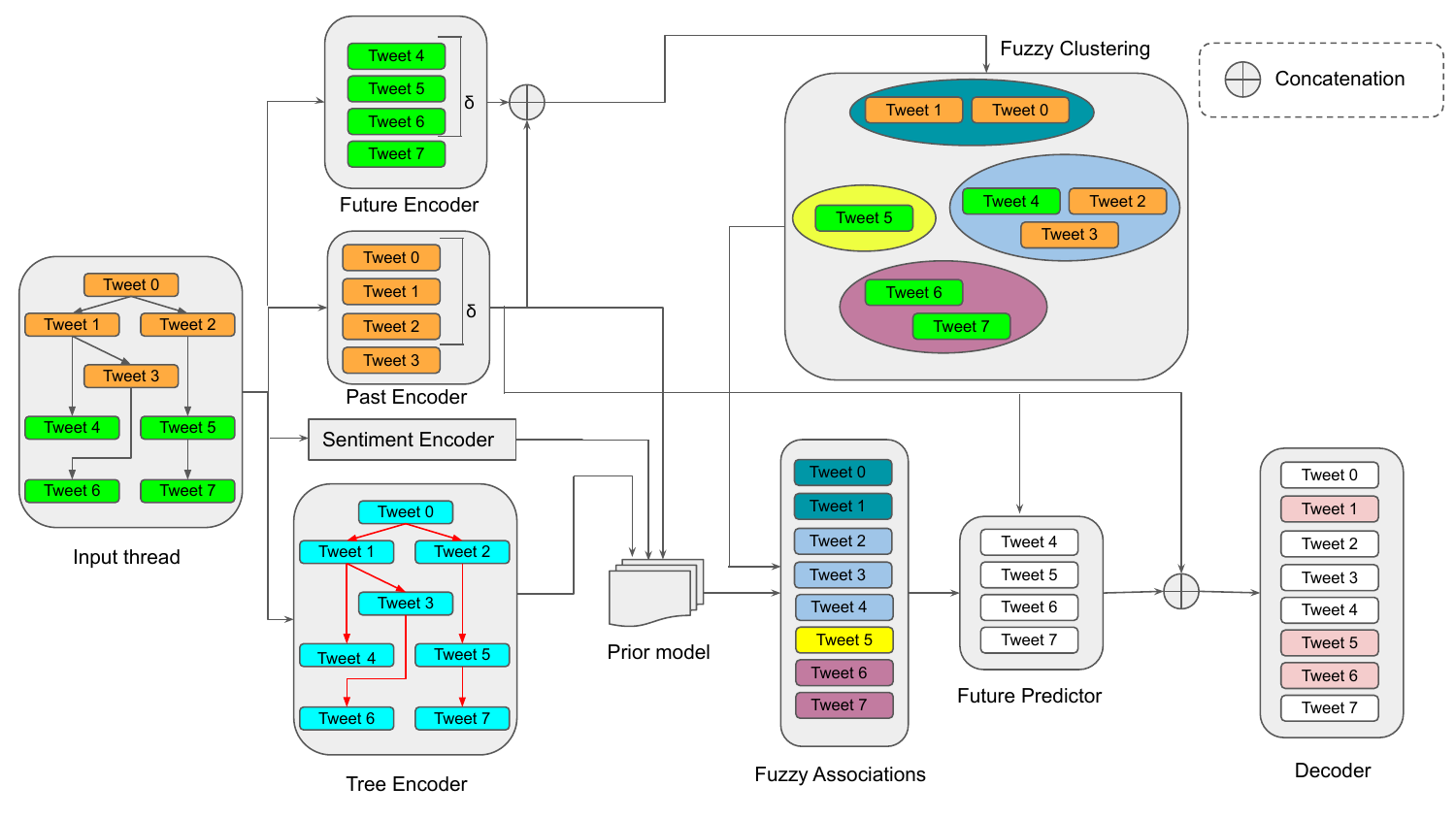}
    \caption{An example processing of an abstract tweet thread using our model. We show the various stages and representations learned by our model and how they combine to profile the final hate intensity of future replies. We differentiate the earlier and future tweets in the input threads using \textcolor{orange}{orange} and \textcolor{green}{green} colors. $\delta$ is the size of the sliding window of tweets. The tree encoder shows differences between the node and edge representations. Fuzzy clustering and fuzzy associations share a common color code to show the assignment of cluster memberships. The final decoder block represents some tweets being assigned high hate intensity scores (i.e., tweets in red)}
    \label{fig:running_example}
    \vspace{5mm}
\end{figure}

\red{Figure \ref{fig:running_example} illustrates an example processing of an abstract tweet thread using our model. Four representations are derived from the input tweet thread: History, Future, Sentiment and Graph representations. The history and future representations are fed into the Fuzzy Clustering algorithm to predict the cluster centers. In parallel, a Prior model is trained on history, sentiment and graph representations forming the prior knowledge. This is combined with cluster centers to learn/assign the cluster membership probabilities in the fuzzy associations step. Further, these probabilities are combined with history representations in the Future Predictor to predict the latent representation of future hate trends. In the final step, the future latent representation is concatenated with the history representation and fed into the decoder to predict the overall hate trend of the entire conversation thread.}

\subsection{Time-series Representative Learning}
\label{sec:time-series}
The vector of the conversation thread $\mathcal{R}_{s_{(1,n)}}$ can be viewed as a collection of irregularly lengthening time series (window-wise hatred intensity profiles). The Dynamic Time Warping (DTW) distance metric and its derivatives are used to group comparable trends together in state-of-the-art approaches for clustering irregular time series \cite{niennattrakul2007clustering}. DTW's precision in mapping time series similarity is beneficial, but the noisy and volatile character of the data points in the present research prevents it from showing high efficiency in clustering comparable hatred patterns into a single stratum. We propose an autoencoder to translate each conversation thread $\mathcal{R}_{(p,q)}$ in $\mathcal{R}_{s_{(1,n)}}$ to a low-dimensional latent representation in order to capture a more suitable representation of the time series. We also propose a multi-encoder strategy instead of a single encoder-decoder design as proposed in \cite{sun2021three}. 

\subsection{Proposed Autoencoder}
\label{sec:autoencoder}
To capture the hate intensity series, the autoencoder module seeks to learn a low-dimensional representation. In {\model}, we specifically construct two encoders and a single decoder. Using the two encoders, the model can learn the representations of historical and future hate intensity sequences separately. The decoder then uses the representations to reconstruct the original hate intensity sequence.

\subsubsection{Encoder}
Several studies provide models for univariate and multivariate \cite{yang2018design, ismail2020inceptiontime} time series to mine the hidden patterns of various sequences. We use a state-of-the-art Inception-Time \cite{ismail2020inceptiontime} module to automatically extract features and represent hate intensity sequences, denoted as: 

\begin{equation}
    \mathcal{X}_m = \mathcal{E}_t(\mathcal{R}_{s_{(1,n)}}),
    \label{equ:time_inception}
\end{equation}
where $\mathcal{X}_m \in \mathbb{R}^{s \times n \times 4}$ is the multivariate intermediate representation of $\mathcal{R}_{s_{(1,n)}}$. The Inception-Time module is denoted by $\mathcal{E}_t(\cdot)$. After the Inception-Time module, we flatten $\mathcal{X}_m$ and use multilayer perceptrons as the classification stage. Finally, the representations are written down as follows: 

\begin{equation}
    \begin{aligned}
    \mathcal{X}_o = \mathcal{E}_{lt}(flatten(\mathcal{X}_m)),
    \end{aligned}
    \label{equ:encoder_structure}
\end{equation}
Here, $\mathcal{X}_m$ is transformed into a one-dimensional vector using $flatten(\cdot)$. $\mathcal{E}_{lt}(\cdot)$ denotes multilayer perceptrons that have been trained to learn the best representation of the hate intensity sequence. 

We develop the history encoder ($\mathcal{E}_h$) and the future encoder ($\mathcal{E}_f$) based on Equations (\ref{equ:time_inception}) and (\ref{equ:encoder_structure}) to encode hate intensity sequences of both historical and future conversation threads: 

\begin{equation}
\begin{aligned}
    \mathcal{X}_h &= \mathcal{E}_h(\mathcal{R}_{s(1,t_h)}) \\
    \mathcal{X}_f &= \mathcal{E}_f(\mathcal{R}_{s(t_{h}+1,t_f)})
\end{aligned}
\end{equation}
where $\mathcal{X}_h \in \mathbb{R}^{s \times N_{\mathcal{X}_h}}$ and $\mathcal{X}_f \in \mathbb{R}^{s \times N_{\mathcal{X}_f}}$ denote the historical and future latent representations, respectively. $N_{\mathcal{X}_h}$ and $N_{\mathcal{X}_f}$ represent the length of $\mathcal{X}_h$ and $\mathcal{X}_f$, respectively.

\subsubsection{Decoder\label{model_decoder}} Although we use two encoders to convert the hate intensity of each conversation thread into two latent representations, $\mathcal{X}_h$ and $\mathcal{X}_f$, we only employ one decoder, $\mathcal{D}(\cdot)$ to return the latent representation to the original input. The decoder is trained to reconstruct the original hate intensity profile per conversation thread by concatenating the two latent representations. The decoder's operation can be summarised as follows: 
\begin{equation}
    \mathcal{R}^*_{s_{(1,n)}} = \mathcal{D}([\mathcal{X}_h \oplus \mathcal{X}_f]),
\end{equation}
where $\mathcal{R}^*_{s_{(1,n)}}$ is the reconstructed hate intensity sequence containing both historical and future hate intensity sequences. 

\subsection{Fuzzy Associations}
The hate intensity profiles of conversation threads in our dataset are noisy, volatile, and lack a discernible pattern, as mentioned in Section \ref{sec:time-series}. Our goal is to group similar profiles using low-dimensional latent representations of data obtained by autoencoder (as explained in Section \ref{sec:autoencoder}). Recent research that supports this technique attests to the validity of deep learning-based models' effectiveness in learning hidden features from time-series data for various applications \cite{cui2016multi, ziat2017spatio}. We apply a clustering strategy over the latent representations to aggregate heterogeneous hate intensity profiles into (near-) homogeneous clusters. Finding meaningful correlations in data is unduly dependent on the number of clusters $j$ and the cluster centers in this unsupervised context. As a result, rather than restricting each profile to a single cluster, we apply a fuzzy clustering approach and use the membership probabilities as a feature embedding.



We define the combined latent space $\mathcal{X}$ as,
\begin{equation}
    \mathcal{X} = \mathcal{X}_h \oplus \mathcal{X}_f.
\end{equation}

Rather than using a hard clustering strategy, in which each profile's association to the nearest cluster is fixed, we use cluster membership probability using a fuzzy clustering approach instead. The membership probability vector, denoted by $P$($C_{c1}, C_{c2}, \dots, C_{cj}$), reflects the associative probabilities of each cluster with the provided chain, where $C_{ci}$ denotes the cluster center of the $i^{th}$ cluster. 

\subsubsection{Fuzzy Clustering \label{fuzzy_clustering}}
We cluster on the combined latent representation $\mathcal{X}$ because our goal is to uncover correlations between each hate intensity profile and homogeneity groupings. To locate the clusters $C_c$, which is the collection of cluster centers, we use a state-of-the-art fuzzy clustering model \cite{reynolds2009gaussian}, indicated by $\mathcal{GM}(\cdot)$: 

\begin{equation}
    C_c = \mathcal{GM}(\mathcal{X}) = (C_{c1}, C_{c2}, \cdots, C_{cj}).
\end{equation}
where $j$ is the pre-defined number of clusters, and $C_{ci}$ is the
cluster centre of the $i^{th}$ cluster.



\begin{figure}
    \centering
    \subfigure[The tree encoder $\mathcal{TE}(\cdot)$]{
    \includegraphics[width=0.7\textwidth]{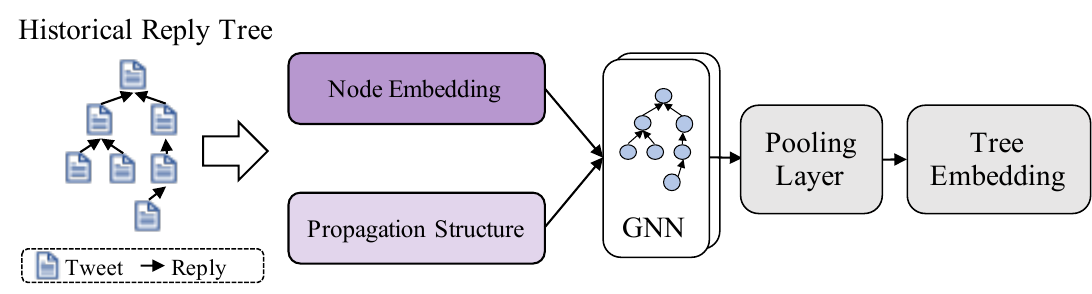}
    \label{subfig:tree_encoder}
    }
    \subfigure[The Node Embedding Module]{
    \includegraphics[width=0.2\textwidth]{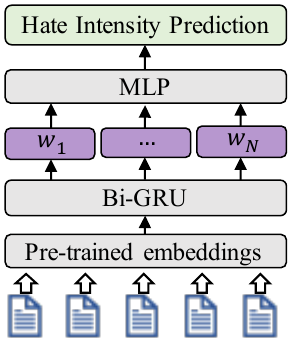}
    \label{subfig:pre_emb}
    }
    \caption{The tree encoder $\mathcal{TE}(\cdot)$ for generating graph-level representations for each conversation thread.}
    \label{fig:tree_encoder}
\end{figure}


\subsection{Tree Encoder} 
Twitter conversation threads typically contain numerous replies (i.e., tweets) with various propagation topological structures. These propagation structures capture the relationships between the tweets; modeling the topological structures would enrich the conversation thread's representation for downstream machine learning tasks~\cite{cao2020popularity, bian2020rumor}. For our hate intensity prediction task, we design a tree encoder $\mathcal{TE(\cdot)}$ to fuse the information of tweet content and the structural information to represent the conversation tree. The tree encoding process is presented in Fig~\ref{subfig:tree_encoder}. Specifically, to learn a conversation thread's representation, we first learn the node embedding by designing a sub-module using a hate intensity prediction task to learn the tweet's representation. Next, we leverage a Graph Neural Network (GNN) \citep{scarselli2008graph} to learn the conversation tree embedding by fusing the conversation tree structural information and the node embedding. 


\textbf{Learning Node Embedding.} We adopt the tweet-level hate intensity prediction task as the supervised signal for learning the embedding of the node (i.e., tweet). We first leverage pre-trained Word2Vec~\cite{mikolov2013distributed} word embedding to represent the words in individual tweets. Next, we use a two-layer bidirectional gated recurrent unit (BiGRU) to learn the individual post's representation, and the tweet-level hate scores are used as training signals. Specifically, the post representations are fed into a two-layer MLP classifier to predict the post's hate label. The model is formulated as follows:


\begin{equation}
    \begin{aligned}{}
    \bm{w}_i &= Bi\text{-}GRU(<w_0,w_1,\cdots,w_N>), \\
    \bm{\hat{p}}_k &= {Att}(<w_0,w_1,\cdots,w_N>), \\
    \hat{y}_k &= {MLP}(\bm{\hat{p}}_k),
    \end{aligned}
    \label{formula_p}
\end{equation}
where $\hat{y}_k$ is the predicted hate score for the $k$ tweet. We can train the model's parameters with the loss function below and the Adam optimizer. We can obtain the tweet-level embedding $\hat{\bm{p}}_k$. The loss function ($\mathcal{L}_h$) is defined as follows:
\begin{equation}
    \mathcal{L}_h = \sum_{k=1}^{N}(y_k-\hat{y}_k)^2 + reg(\Theta)
\end{equation}
where $N$ is the number of  tweets in all conversation threads, $reg(\cdot)$ is a regularization function to alleviate overfitting, and $\Theta$ is a parameter for the fine-tuning model.

\textbf{Learning Tree Embedding.} The goal of learning the tree embedding of the conversation thread is to leverage useful propagation structural information to perform early hate intensity prediction. Working towards this goal, we leverage GNNs to model the tree structure in the conversation threads. Specifically, GNNs model the conversation thread structural information as a directed graph with the node represented using the node embeddings learned from the previous section: 

\begin{equation}
    \bm{P}^{(l)} = \sigma(\bm{D}^{-1/2}\bm{A}\bm{D}^{-1/2}\bm{P}^{(l-1)}\bm{W}^{(l-1)})
\end{equation}
where $\bm{W}$ is the trainable parameter, $\bm{P}^{(l-1)} \in \mathbb{R}^{n \times d}$ is the representation of all tweets, $\bm{P}^{0}$ is initialized by all tweet-level embeddings $\hat{\bm{p}}_{k}, k \in \{1,2,\cdots,t_h\}$ learned using formula (\ref{formula_p}), $\bm{A}$ and $\bm{D}$ are the adjacency matrix and degree matrix of the conversation thread tree structure, respectively. Finally, using the average pooling operation, we can obtain the graph-level embedding of the reply tree, denoted $\bm{\mathcal{X}}_{hs}$.

\subsection{Boosting Prediction with Prior Knowledge} 
The task of predicting the hate intensity of upcoming replies, provided limited history $\mathcal{R}^*_{s_{(1,t_h)}}$, is strenuous even for state-of-the-art deep learning models due to the noisy, volatile, and heterogeneous nature of the time-series hate intensity profiles. To address this, we introduce the notion of prior knowledge to the prediction component of our pipeline as the weighted sum of the cluster centers, where the weights correspond to the cluster membership probabilities for the new chain, denoted by $\mathcal{P}^*(C_{c1}, C_{c2}, \dots, C_{cj}$). We define prior knowledge as follows:
\begin{equation}
    \begin{aligned}
    \mathcal{X}^{c} &= \sum_{i \in (0 \leq i \leq j)} C_{ci} P^*(C_{ci}) \\
    \end{aligned}
\end{equation}
Note that $C_c$ is calculated over $\mathcal{X}$, i.e., the combined latent representation. Therefore, to calculate the complete membership probability vector for a new chain, we cannot directly use the fuzzy clustering model $\mathcal{GM}(\cdot)$. Instead, we construct a prior model $\mathcal{PR}(\cdot)$ to predict the membership probabilities for new chains using only the latent representation of the history $\mathcal{X}_h$, the sentiment feature $\mathcal{S}_{s(1,t_h)}$, the historical replies $\mathcal{R}_{s_{(1,t_h)}}$ and propagation tree structure $G_{\varphi}^{(1,t_h)}$.


\begin{equation}
    \mathcal{PR}(\mathcal{E}_{h}(\mathcal{R}_{s_{(1,t_h)}}), \mathcal{S}_{s_{(1,t_h)}},\mathcal{TE}(\mathcal{R}_{s_{(1,t_h)}}, G_{\varphi}^{(1,t_h)})) = \mathcal{P}^*(C_{c1},C_{c2},\cdots,C_{cj})
\end{equation}
The precision of the predictions by the prior regression
model is measured by comparing $\mathcal{P}^*(C_{c1}, C_{c2}, \dots, C_{cj}$) against $\mathcal{P}(C_{c1}, C_{c2}, \dots, C_{cj}$).

\subsubsection{Estimating Latent Representation of Upcoming Conversation Threads}
The designed decoder needs the latent representations of historical and future conversation threads to reconstruct the complete hate intensity profile. However, our primary objective is to predict the future hate intensity profile based on historical information. To provide the decoder with the ability to perform prediction, we propose a future representation predictor that uses historical information to estimate the latent representation of future hate intensity profile $\mathcal{X}_f$. Then, combined with the latent representation of historical conversation threads, the output can be fed into the decoder. 

Specifically, we utilize the prior knowledge extracted by the fuzzy clustering module and the latent representation $\mathcal{X}_c$ of the historical conversation threads. To avoid the estimation problem being unduly influenced by prior knowledge, we design the predictor in two steps. In the first step, for each conversation thread, the prior latent representation of historical conversation threads $X_h^c$ contains the required information to predict the latent representation of future conversation threads. Moreover, we have the expected latent representation $\mathcal{X}_h$ encoded from the initial historical replay chains. As a result, we use the difference operator on the expected ($\mathcal{X}_h$) and estimated priors ($\mathcal{X}^c_h$) of the historical conversation threads to evaluate the deviation of these two representations. Secondly, a single-layer perceptron $\mathcal{FP}_d(\cdot)$ is used to learn the representation ($X_d$) in a hidden space, indicating the dissimilarity between the prior knowledge and the history conversation threads. 
\begin{equation}
\begin{aligned}
    \mathcal{X}_s &= \mathcal{X}_h \ominus \mathcal{X}^c_h \\
    \mathcal{X}_d &= \mathcal{FP}_d(\mathcal{X}_s)
\end{aligned}
\end{equation}


We finally obtain the $X_{hc}$ vector by concatenating the provided input $X_h$, the pre-processed prior $X_d$ and $X^c_f$ as, $X_{hc} = X_h \oplus X_d \oplus X^c_f$. The second stage is the deep linear transformation model $\mathcal{FP}_p(\cdot)$ that predicts the upcoming hate intensity in the latent space $X^*_f$ as follows:

\begin{equation}
   \mathcal{X}^{*}_f = \mathcal{FP}_p(\mathcal{X}_{hc})
\end{equation}

\subsection{Decoding the Future}
We use the decoder module trained in Section \ref{model_decoder} to forecast entire hate intensity profiles (i.e., $\mathcal{R}^*_{s(1,n)}$) based on the expected latent representation of the upcoming hate intensity $\mathcal{X}^*_f$. In particular, we concatenate the original latent representation ($\mathcal{X}_h$) of historical hate intensity sequences with the expected future representation ($\mathcal{X}^*_f$), denoted as: 
\begin{equation}
    \mathcal{X}^* = \mathcal{X}_h \oplus \mathcal{X}^*_f
\end{equation}
where $\mathcal{X}^*$ is the predicted hate intensity profile of the upcoming conversation thread in the latent space. The decoder is designed to be a mirror of the encoder. The decoder is defined as follows: 

\begin{equation}
    \mathcal{R}^*_{s_{(1,n)}} = \mathcal{D}(\mathcal{X}^*)
\end{equation}
where $\mathcal{R}^*_{s_{(1,n)}}$ denotes the expected hate intensity profile for the next future conversation thread. Finally, we compare the predicted and original hate intensity score sequences (i.e., $\mathcal{R}_{s_{(1,n)}}$) to evaluate and report the performance on several metrics.

\subsection{Comparison with {DRAGNET}}
From the modeling standpoint, \model\ has advanced DRAGNET~\cite{sahnan2021better} on two aspects: \model\ exploits the tweet-level semantics and the conversation-level structural information to improve hate intensity prediction. DRAGNET leveraged sentiment features, calculated as each tweet's similarity in the conversation threads to the root tweet. However, this approach largely ignores the complex relationships among the tweets in the conversation threads. Furthermore, as illustrated in our earlier example in Fig.~\ref{fig:prop_example}, the root tweet might be benign, and anchoring the sentiment features computation base on the root tweet may not provide an accurate forecast of the sentiment of the subsequent tweets in the conversation threads. Therefore, \model\ addressed this limitation by modeling the tweets' semantics with an additional hate intensity prediction task at the tweet level and the conversation thread structure with GNNs as the tree encoder. The intuition is that by learning the node (i.e., tweets) and tree (i.e., conversation thread) representations using GNNs, we are able to capture the dependency among the tweets in the conversation threads and extract unique structural patterns that could improve the prediction of hate intensity for conversation thread.

\section{Datasets}
\label{sec:dataset}
We evaluate \model\ on three publicly available large Twitter datasets. These datasets contain a large amount of Twitter conversation threads, which were originally collected for other hate speech-related studies. Table ~\ref{tab:dataset_stats} shows the statistics of the datasets.

\begin{table*}[ht]
\centering
\resizebox{\textwidth}{!}{\begin{tabular}{l|c|c|c|c|c|c}
\hline
\multirow{2}{*}{\textbf{Dataset}} &\multirow{2}{*}{\textbf{\#Conversation threads}} & \multicolumn{3}{c|}{\textbf{Conversation thread length}} & \multirow{2}{*}{\textbf{\#Tweets}} & \multirow{2}{*}{\textbf{\#Unique users}} \\
\cline{3-5}
& & \multicolumn{1}{c|}{\textbf{Min.}} & \multicolumn{1}{c|}{\textbf{Max.}} & \multicolumn{1}{c|}{\textbf{Avg.}} &  & \\
\hline \hline
Anti-Racism & 3,500 & 1 & 582 & 200 & 750,235 & 620,437\\ \hline
Anti-Social & 668,082 & 1 & 20014 & 31 & 40,385,257 & 4,980,160\\ \hline
Anti-Asian & 218,790 & 1 & 2822 & 35 & 206,348,565 & 23,895,911\\ \hline
\end{tabular}}%
\caption{The statistics of the three datasets used in our experiments.}
\label{tab:dataset_stats}
\vspace{5mm}
\end{table*}

In the \emph{Anti-Racism} dataset \citep{sahnan2021better}, the authors manually identified various real-world events using a hashtag-based matching via the Twitter API. During the tumultuous year of 2020, many topics polarised the discussion, such as the 2020 US Presidential election, the Brexit referendum in the UK, and extending similar political issues across the US, the UK, and India. Adding to the diversity of the dataset, \emph{sinophobic} tweets attributing coronavirus to China are also curated in the dataset, with most mentions (in tweets) as ``China virus''. The final dataset comprises ~3500 Twitter conversation threads with over 750K tweets.

In the \emph{Anti-Social} dataset \citep{awal_analyzing_2020}, the COVID-19-related tweets were collected from the Twitter platform. The authors performed query searches using case-sensitive keywords such as ``covid-19'', ``COVID-19'', ``Coronavirus'', ``coronavirus'' and ``corona''. Leveraging the Twitter Streaming API, the authors collected the conversations related to these queried tweets. Using this approach, the authors collected over $650K$ Twitter conversations with over $40M$ tweets published between March 17 and 2020 to April 28, 2020.

In the \emph{Anti-Asian} dataset \citep{he_racism_2021}, the authors focused on the Asian hate speech surrounding the COVID-19 discussions. They followed a similar keyword-based approach but as a two-step process. The \emph{covid-19 keywords} are used to narrow the COVID-19-related tweets, following which \emph{hate keywords} about anti-Asian hate is used to segregate the tweets as defined by the collection for the dataset. Finally, \emph{counter-speech keywords} are keywords and hashtags used to counter hate speech and support Asians. It comprises of 42 keywords. The authors used a combination of Twitter Streaming API and Twitter Search API to collect real-time tweets between January 15, 2020 and March 26, 2021. The dataset comprises over $210K$ Twitter conversation threads with more than ~$206M$ tweets. 

As mentioned in Section \ref{sec:preliminaries}, our main task is building hate-intensity profiles for each dataset's tweets. However, it is important to note that we do not use the class labels in the datasets for our task as they do not contain ground truth for the hate intensity of tweets. Instead, we will be using the tweet text to derive the hate intensity scores (as discussed in Section \ref{sec:preliminaries}).



\begin{figure}[ht]
\subfigure[Anti-Racism]{\includegraphics[width=0.45\textwidth]{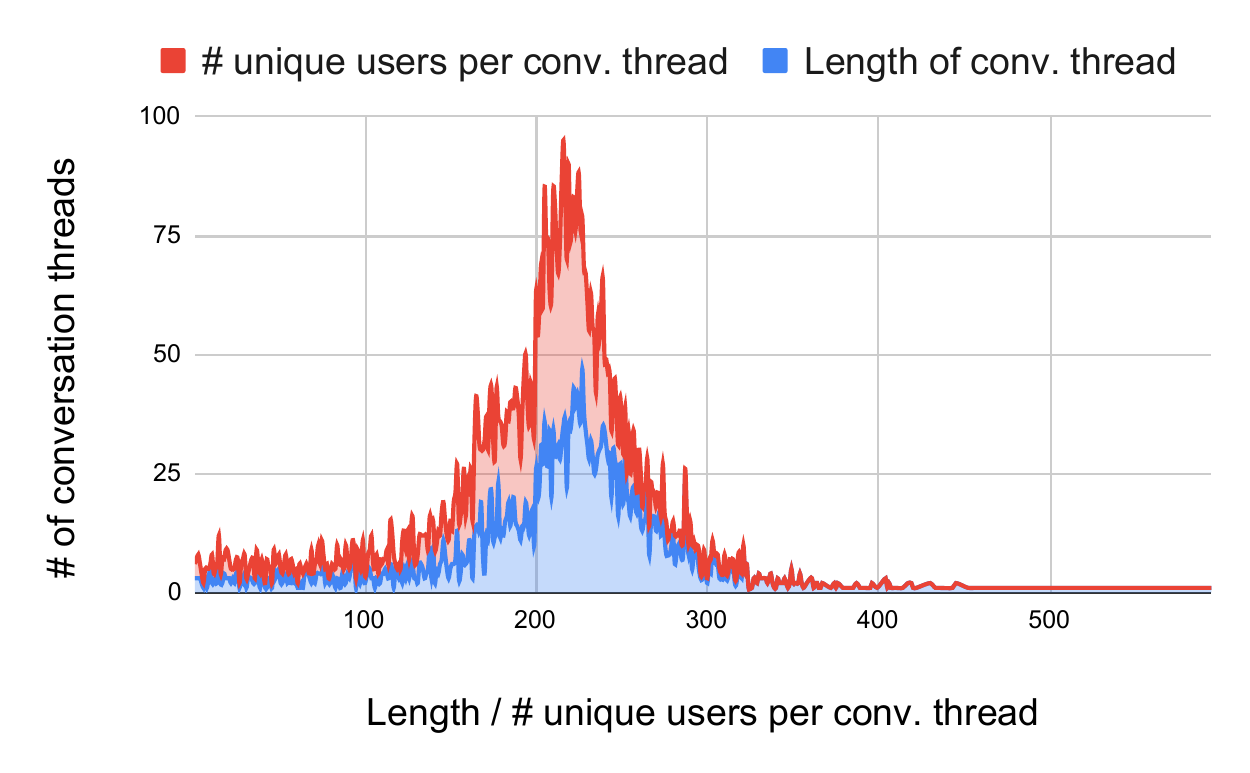}}
\qquad
\subfigure[Anti-Social]{\includegraphics[width=0.45\textwidth]{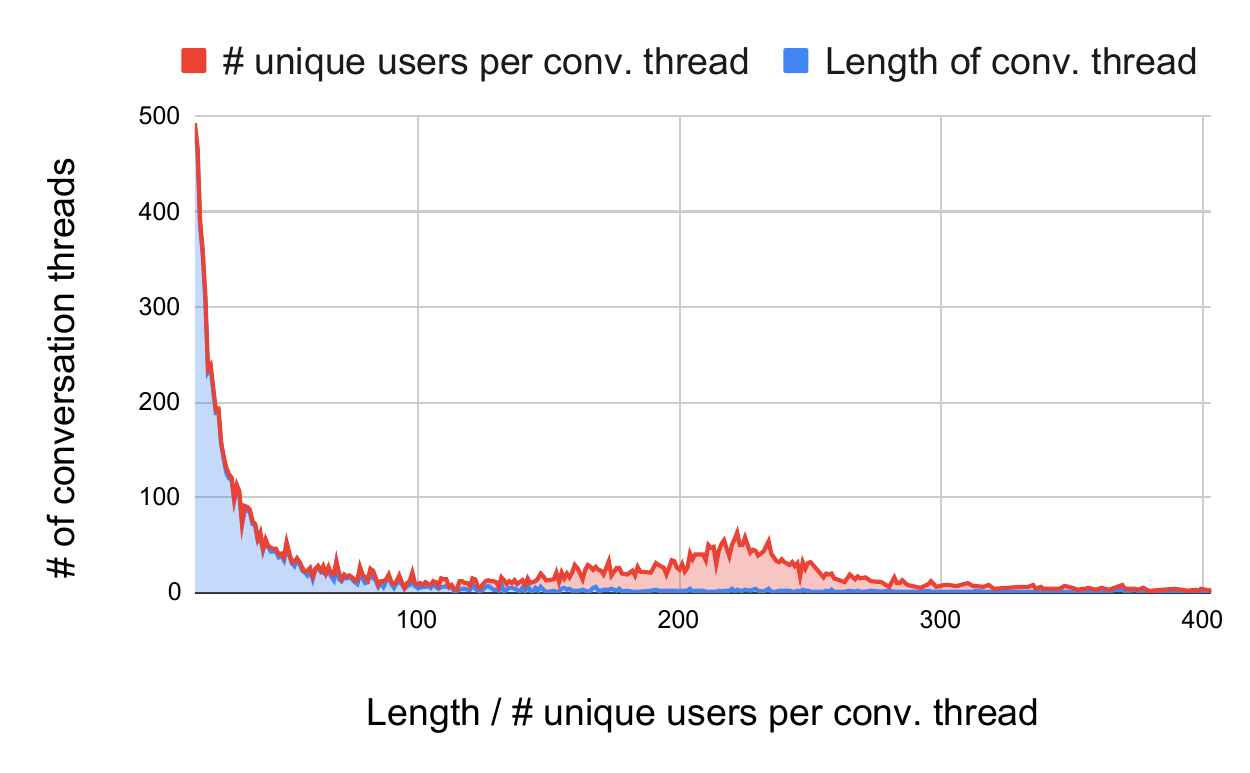}}
\\
\centering\subfigure[Anti-Asian]{\includegraphics[width=0.45\textwidth]{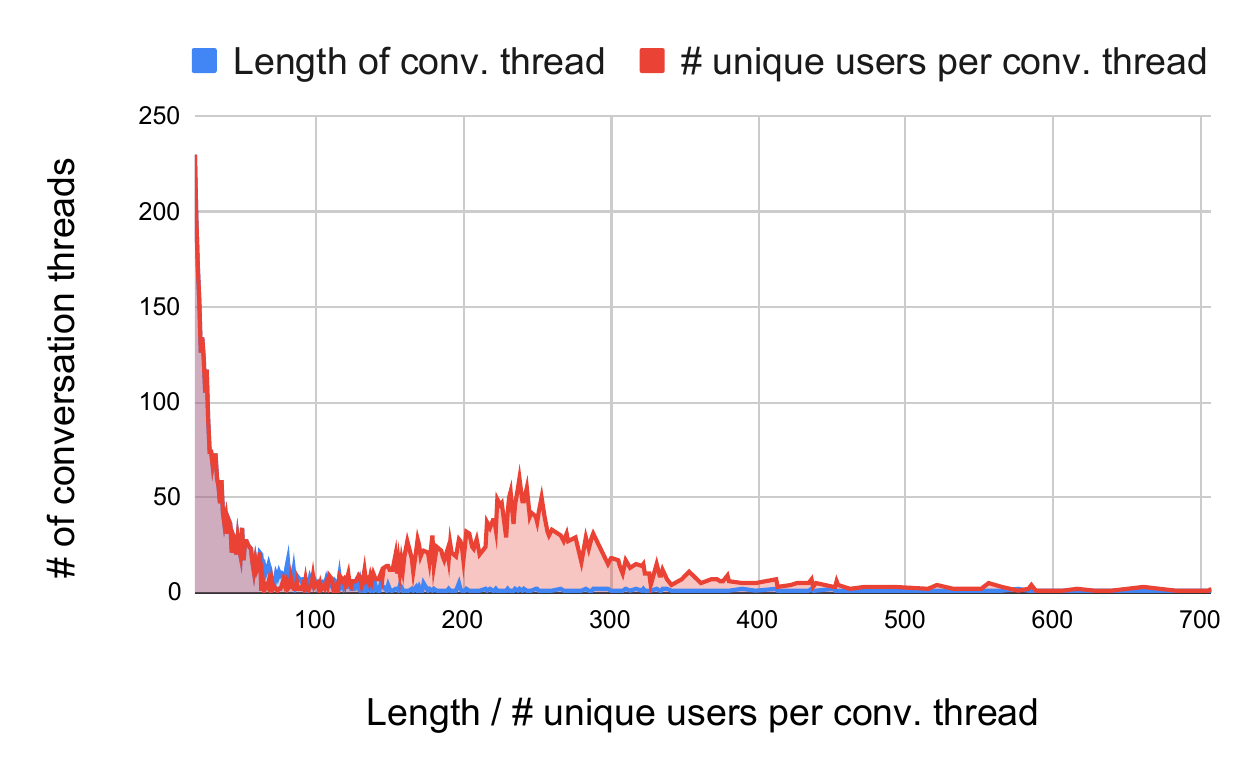}}
\caption{Distribution of number of conversation threads vs length of each conversation thread and the number of unique users in respective conversation threads for all three datasets. Note that for the \emph{Anti-Social} and \emph{Anti-Asian} datasets, the x-axis ticks start from 10 and ends earlier than the maximum length as described in Table ~\ref{tab:dataset_stats} to show the distribution more clearly.}
\label{fig:data_stats}
\vspace{5mm}
\end{figure}

\paragraph{\bf Length of Conversation Threads} Figure ~\ref{fig:data_stats}(a) shows the length of conversation threads over the number of conversation threads for the \emph{Anti-Racism} dataset. The average conversation thread length is around $200$, with the maximum length being $582$. On the other hand, Figures ~\ref{fig:data_stats}(b) and \ref{fig:data_stats}(c) show a clipped version of the entire graph for better visualization of the distribution. Most of the tweets constitute lengths less than or equal to $100$. To understand the full picture, the numerical statistics of \emph{Anti-Social} and \emph{Anti-Asian} are represented in Table ~\ref{tab:dataset_stats}. The maximum length in the \emph{Anti-Social} dataset goes as high as $20014$. However, the average length of the dataset is around 31. Along similar lines, the \emph{Anti-Asian} dataset also touches a maximum conversation thread length of $2822$ with an average length of $35$ across the tweet threads. 

\paragraph{\bf Number of Unique Users} Figure ~\ref{fig:data_stats} also highlights the number of unique users per reply thread. For the \emph{Anti-Racism} dataset, it can be observed that the unique users and the lengths follow a similar distribution. The increased number of unique users can be attributed to limited re-engagement by the same users for a particular thread. Furthermore, both the \emph{Anti-Social} and \emph{Anti-Asian} datasets exhibit a high variation in the length of the threads with a consistent presence of threads in almost all lengths. Correspondingly, the unique users touch about $65$ for about 225 threads in the case of \emph{Anti-Social}, whereas $70$ unique users for about $240$ threads in \emph{Anti-Asian}.

\paragraph{\bf Number of Conversation threads and Tweets} The conversation threads and the total number of tweets are presented in Table ~\ref{tab:dataset_stats}. Compared to the \emph{Anti-Racism} dataset, the other two datasets possess a vast number of conversation threads, forming the ideal datasets to test the performance of the models. However, the number of tweets is ~$206M$ for \emph{Anti-Asian}, whose conversation threads are less in number than \emph{Anti-Social}, which has about ~$40M$ tweets in total. This shows the variation in conversation thread length for different datasets and the corresponding unique users overall, adding diversity to the datasets.

\section{Experiments}
\label{sec:experiment}


\subsection{Baselines\label{sec:baselines}}
There are few studies that predict hate intensity profiles in Twitter conversations~\citep{sahnan2021better,dahiya_would_2021}. As a result, we use time-series forecasting and temporal pattern model based on both conventional and deep learning as baselines. 

\begin{itemize}
    \item \textbf{LSTM~\citep{elsworth_time_2020}:} Long short-term memory is a neural network architecture with feedback connections that is very effective. LSTMs, unlike RNNs, perform better on long-range predictions, such as time-series problems. As suggested by \citet{elsworth_time_2020}, we employ a stacked LSTM. We use ReLU as the activation function for each layer to forecast the hate intensity profile. 
    \item \textbf{CNN~\citep{mehtab_robust_2020}:} A convolutional neural network considers the hate intensities' time information to better profile the tweet's upcoming replies. We employ a 1-D CNN architecture with ReLU activation as the last layer. The kernel size is 2, and the number of filters employed is 64. 
    \item \textbf{N-Beats~\citep{oreshkin_n-beats_2020}:} \textbf{N}eural \textbf{B}asis \textbf{E}xpansion \textbf{A}nalysis for interpretable \textbf{T}ime \textbf{S}eries forecasting is a deep learning model designed to tackle the univariate time-series problem. With a very deep stack of fully-connected layers, it incorporates forward and backward residual linkages. 
    \item \textbf{DeepAR~\citep{salinas2020deepar}: } It is a supervised learning approach for forecasting scalar time-series using Recurrent Neural Networks (RNN), which is achieved by auto-regressively training the RNN on numerous related time-series data. 
    \item \textbf{TFT~\citep{lim_temporal_2020}:} {\bf T}emporal {\bf F}usion {\bf T}ransformers are used in conjunction with multi-horizon forecasting to help the model mix known future inputs and extract exogenous correlations from the time series' past input data. The model employs a self-attention-based architecture to give interpretable time-series insights. 
    \item \textbf{ForGAN~\citep{koochali_probabilistic_2019}:} Probabilistic forecasting of sensory data using generative adversarial networks employs an adversarial network to learn data generating distributions and compute probabilistic forecasts over them. In a time-series forecasting problem, the model claims to learn the conditional probability distribution of future values with no quantile crossing or reliance on the prior distribution.  
    \item \textbf{DRAGNET~\citep{sahnan2021better}}: It is a deep stratified learning approach. To create the latent representation of past and future tweets from the specified point of reference as hyperparameter, an autoencoder is utilized using the Inception-Time module \citep{ismail2020inceptiontime}. This is combined with a fuzzy clustering algorithm to forecast the future hate intensity sequence, which is reinforced with sentiment features to determine the correlation of each tweet with the root tweet. 
\end{itemize}


\subsection{Experimental Setup\label{sec:experimental_analysis}}
The hyperparameters used for all the experiments are as follows: $\delta$ = 10, $w = 0.6$, $t_h = 25$, $t_f = 275$, $n = 300$, $j = 15$, $N_{X_h} = 32$, $N_{X_f} = 128$. In the auto-encoder step, \model\ is implemented with the Inception-Time module \citep{ismail2020inceptiontime} for transformation with variable kernel sizes -- 5,7, and 9.
Sentiment characteristics, history encodings from the auto-encoder, and graph structure information are fed into fully-connected layers in the prior model.
A Gaussian mixture model with full covariance is utilized for fuzzy clustering. Davidson model \citep{davidson_automated_2017} is the default hate-speech classifier. We utilize the \emph{Adam} optimizer with an \emph{lr} = $0.001$ for faster convergence, and the training and test split is set at $80:20$. The train-test split is based on the number of conversation threads since we would like to train and test on complete tweet propagation. 

Our model predicts the future hate intensity sequence, given the tweets and historical hate intensity sequence for the chosen window length. To evaluate, three metrics for this task are used: (i) \textbf{Pearson correlation coefficient} (PCC), where a higher value is better, (ii) \textbf{Root Mean Square Error} (RMSE), and (iii) \textbf{Mean Forecast Error} (MFE), where lower values are better.  

\subsection{Experimental Results and Analysis}
The overall performance of \model\ and baselines on the three datasets are shown in Table~\ref{tab:hate_3model_results}.
Across all datasets, we find that \model\ outperforms the baselines. Specifically, for the \emph{Anti-Racism} dataset, \model\ outperforms the best baseline (N-Beats) by 40\% on PCC. Other assessment parameters show a similar trend, with a 100\%  reduction in RMSE and a $9\times$ reduction in MFE compared to N-Beats. On the contrary, as compared to the baselines on the \emph{Anti-Asian} dataset, both {DRAGNET} and \model\ perform poorly on PCC. One possible explanation is the dataset's restricted chain of responses. \model\, on the other hand, outperforms N-Beats by more than $3\times$ on RMSE when optimizing the inaccuracies on hate intensity profiles.
TFT has a higher MFE score than \emph{DRAGNET}, but \model\ has the highest total score. Along with LSTM and CNN, ForGAN appears to be among the worst performers. On the \emph{Anti-Social} dataset, \model\ outperforms the best baseline, TFT, by 0.223, 0.476, and 0.202, respectively, on PCC, RMSE, and MFE. 
When DRAGNET and \model\ are directly compared, the latter performs better on all three datasets and on all three metrics. This demonstrates how graph information inherent in tweet threads can help forecast hate intensity profiles more accurately. The greatest significant improvement over RMSE is $22.72\%$ on the \emph{Anti-Social} dataset, and the highest increase on PCC is $0.066$ points on the \emph{Anti-Racism} dataset. On the MFE (\emph{Anti-Racism} dataset), DRAGNET performs the best, which can be attributed to customized windowing strategies that reduce forecast error. {DRAGNET}'s improved performance does not apply to other datasets, where \model\ readily outperforms.

\begin{table}[!t]

\centering
\resizebox{\textwidth}{!}{\begin{tabular}{l|c|c|c|c|c|c|c|c|c}
\hline
\multicolumn{1}{c|}{\multirow{2}{*}{\textbf{Model}}} & \multicolumn{3}{c|}{\textbf{Anti-Racism}}                                              & \multicolumn{3}{c|}{\textbf{Anti-Social}}                                              & \multicolumn{3}{c}{\textbf{Anti-Asian}}                                               \\ \cline{2-10} 
\multicolumn{1}{c|}{}                       & \multicolumn{1}{c|}{\textbf{PCC}} & \multicolumn{1}{c|}{\textbf{RMSE}} & \multicolumn{1}{c|}{\textbf{MFE}} & \multicolumn{1}{c|}{\textbf{PCC}} & \multicolumn{1}{c|}{\textbf{\textbf{RMSE}}} & \multicolumn{1}{c|}{\textbf{MFE}} & \multicolumn{1}{c|}{\textbf{PCC}} & \multicolumn{1}{c|}{\textbf{\textbf{RMSE}}} & \multicolumn{1}{c}{\textbf{MFE}} \\ \hline\hline
LSTM & 0.145 & 0.611 & 0.500 & 0.160 & 0.511 & 0.315 & 0.680 & 0.722 & 0.692\\ \hline
CNN & 0.105 & 0.644 & 0.509 & 0.112 & 0.542 & 0.320 & 0.675 & 0.731 & 0.699\\ \hline
DeepAR & 0.310 & 0.484 & 0.065 & 0.180 & 0.490 & 0.275 & 0.682 & 0.748 & 0.708\\ \hline
TFT & 0.469 &  0.437 &  0.076 & 0.376 &  0.630 &  0.333 & 0.562 & 0.866 & 0.125\\ \hline
N-Beats & 0.380 &  0.544 &  0.085 & 0.340 &  0.633 &  0.271 & \bf 0.712 &  0.462 &  0.173\\ \hline
ForGAN & 0.240 &  0.603 &  0.360 & 0.172 &  0.897 &  0.785 & 0.563 & 0.871 & 0.475\\ \hline
DRAGNET & 0.563 &  0.247 &  0.010 & 0.559 &  0.189 &  0.156 & 0.603 &  0.165 &  0.134\\ \hline
\model & \bf 0.629 &  \bf 0.218 &  \bf 0.008 & \bf 0.599  &  \bf 0.154  &  \bf 0.131  & 0.641 &  \bf 0.147 &  \bf 0.116\\ \hline

\end{tabular}}
\caption{Overall performance of \model\ and baselines on the three datasets. The best results are \textbf{bold}.}
\label{tab:hate_3model_results}
\end{table}

\subsection{Ablation Study}
\begin{figure}[ht]
    \centering
    \includegraphics[width=\textwidth]{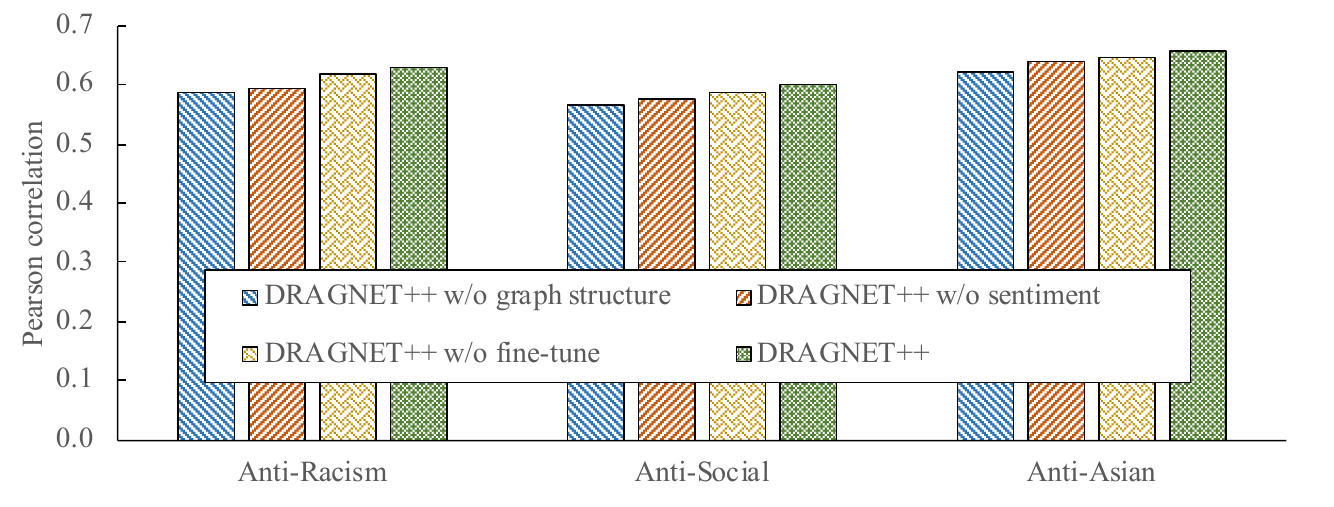}
    \caption{Ablation study: The Effectiveness of components in prior model $\mathcal{PR}(\cdot)$ on three datasets.}
    \label{fig:abaltion_exp}
\end{figure}

The prior model $\mathcal{PR}(\cdot)$ in \model\ uses the propagation graph structure, sentiment information, and fine-tuning operation to improve the representation of historical discussion threads and establish a link between historical and future data. To assess the performance of each component, we remove it one at a time and provide three variants: {\model} w/o graph structure, {\model} w/o sentiment, and {\model} w/o fine-tune. Then, using three datasets, we run experiments to see how the alternative models perform in the end, and the findings are displayed in Fig \ref{fig:abaltion_exp}. 

The model's performance degrades for each of the versions, as seen in Fig \ref{fig:abaltion_exp}. It indicates that each component contributes differently to the prediction of conversation thread labels, resulting in improved hate intensity prediction results. Across all three datasets, {\model} w/o graph structure performs the worst among the four variations. The diffusion patterns and informative material are present in the propagation graph structures of conversation threads; therefore, this is reasonable. Capturing earlier responses and projecting future patterns would help the data. When comparing {\model} w/o sentiment to {\model}, the performance drops slightly, but not as much as when comparing {\model} w/o graph structure. A possible reason could be that the tweet's sentiment is assessed in relation to the root tweet. On the other hand, structural graph information is a more advanced method of recording hate intensity distribution. Furthermore, {\model} outperforms {\model} w/o fine-tuning. It suggests that using the fine-tune technique, word embedding can precisely capture hate-intensity information.

\subsection{Parameter Analysis}
We further analyze the parameters of \model\ to understand its superiority and limitations better. Throughout this study, we compare \model\ and DRAGNET.

\subsubsection{The impact of different window sizes}
\begin{figure}[h]
\subfigure[Anti-Racism]{\includegraphics[width=0.28\textwidth]{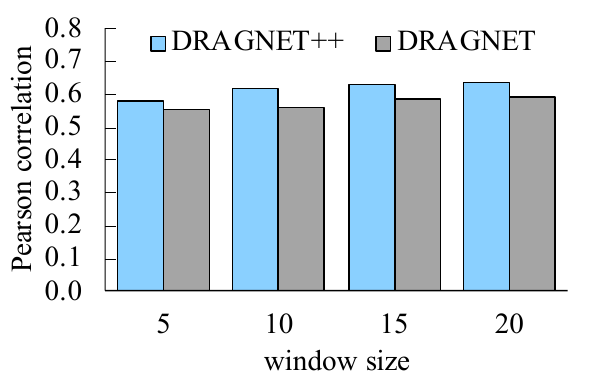}}
\qquad
\subfigure[Anti-Social]{\includegraphics[width=0.28\textwidth]{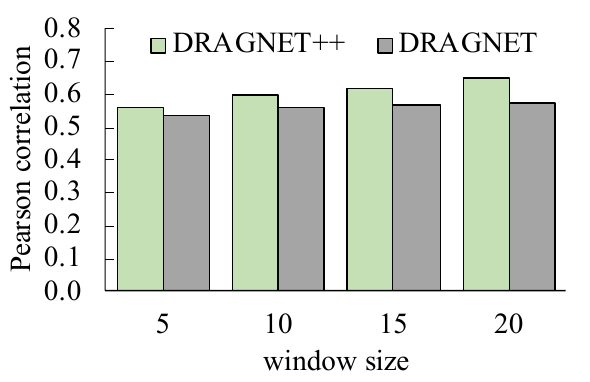}}
\qquad
\subfigure[Anti-Asian]{\includegraphics[width=0.28\textwidth]{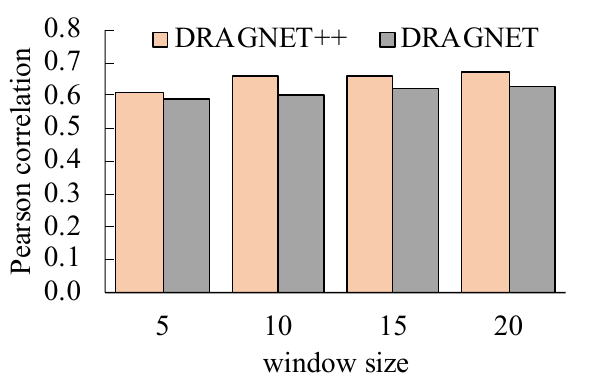}}
\caption{Performance comparison based on window size ($\delta$) of hate intensity profiles.}
\label{fig:window_size}
\end{figure}

The window size utilised in the average rolling process is represented by the hyperparameter $\delta$.
It is possible that the smaller window size would result in more pronounced hate intensity profiles.
We chose the window size in $\{5,10,15,20\}$ to analyse the influence of different window sizes, and the experimental findings are displayed in Fig \ref{fig:window_size}. 
Across all three datasets, we observe that increasing the value of $\delta$ improves the performance of both \model\ and {DRAGNET}. A bigger window size ($\delta$) smooths the values in hate intensity sequences, making it easier for the models to learn the hidden hate intensity pattern. Furthermore, \model\ consistently outperforms {DRAGNET} across all window size settings, demonstrating its efficacy. 


\subsubsection{The impact of history sizes} 
\begin{figure}[h]
\subfigure[Anti-Racism]{\includegraphics[width=0.28\textwidth]{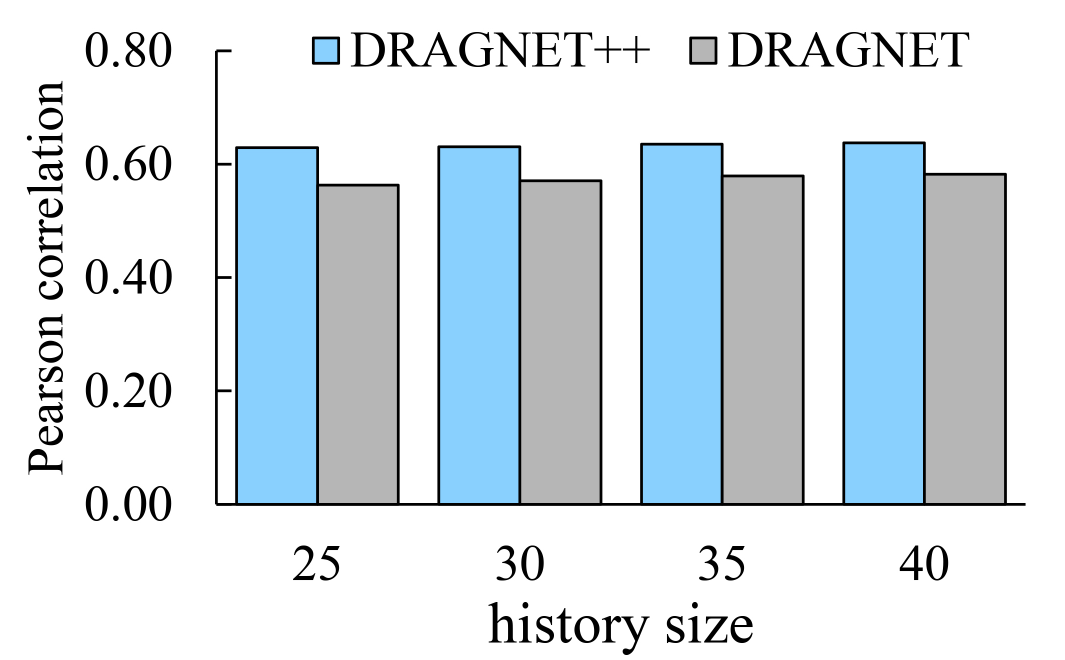}}
\qquad
\subfigure[Anti-Social]{\includegraphics[width=0.28\textwidth]{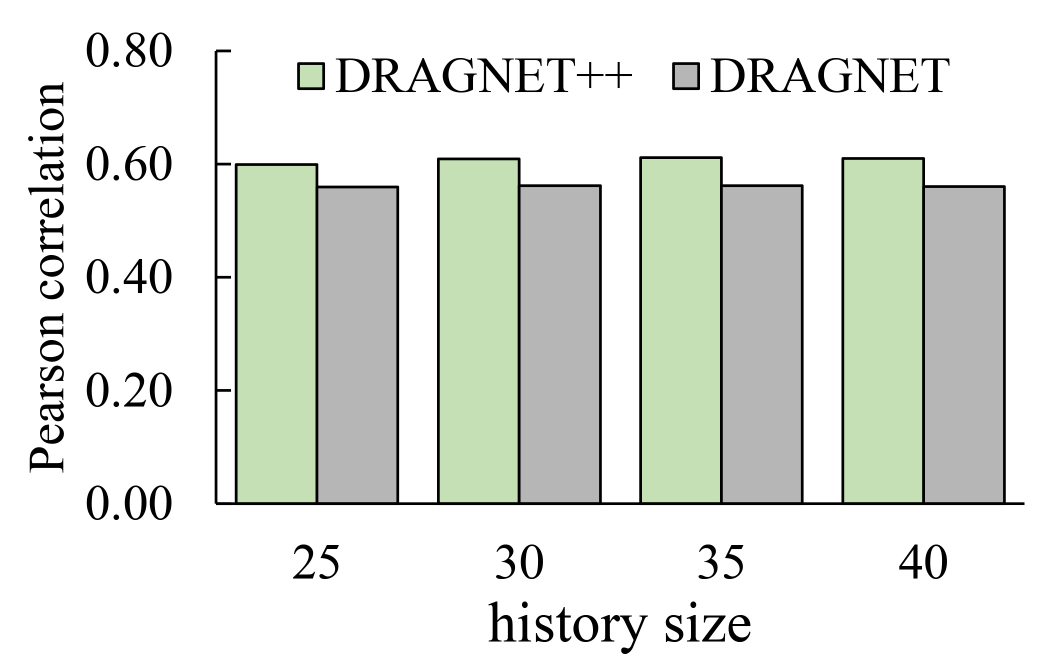}}
\qquad
\subfigure[Anti-Asian]{\includegraphics[width=0.28\textwidth]{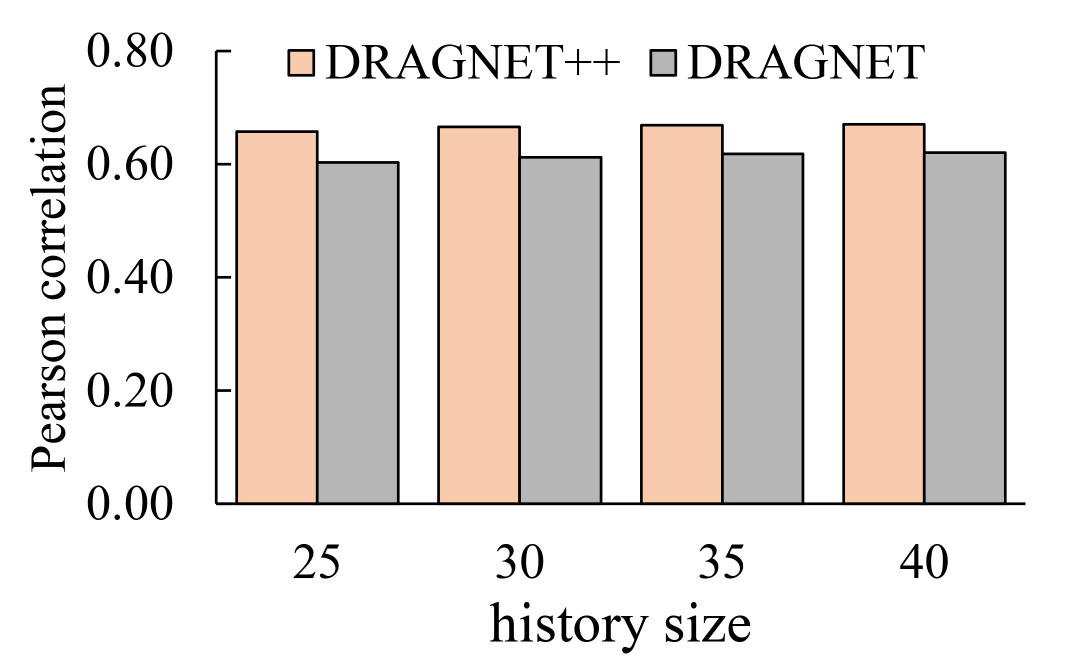}}
\caption{Performance of {\model} on three datasets based on changes to tweet history length at input.}
\label{fig:history_size}
\end{figure}

The beginning history size $t_h$ is the data needed to estimate the whole hate intensity profile for a new conversation thread. The value of the Pearson correlation coefficient for {\model} shows relatively little change as $t_h$ increases, as seen in Fig \ref{fig:history_size}. We can observe that \model\ outperforms  DRAGNET. The model's capacity to generate early predictions decreases as the number of initial responses submitted to the model grows. As a result, in the experiments, we used $t_h$=25 to predict hate intensity. 

\subsubsection{The impact of different numbers of clusters} 
\begin{figure}[h]
\subfigure[Anti-Racism]{\includegraphics[width=0.28\textwidth]{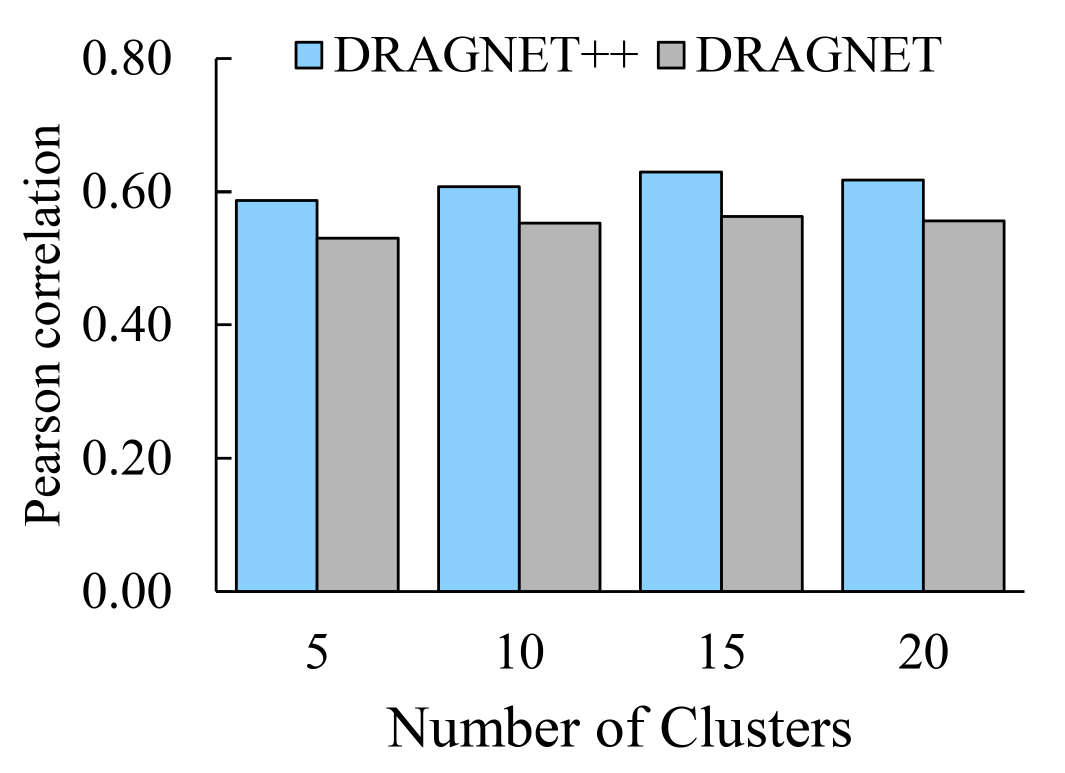}}
\qquad
\subfigure[Anti-Social]{\includegraphics[width=0.28\textwidth]{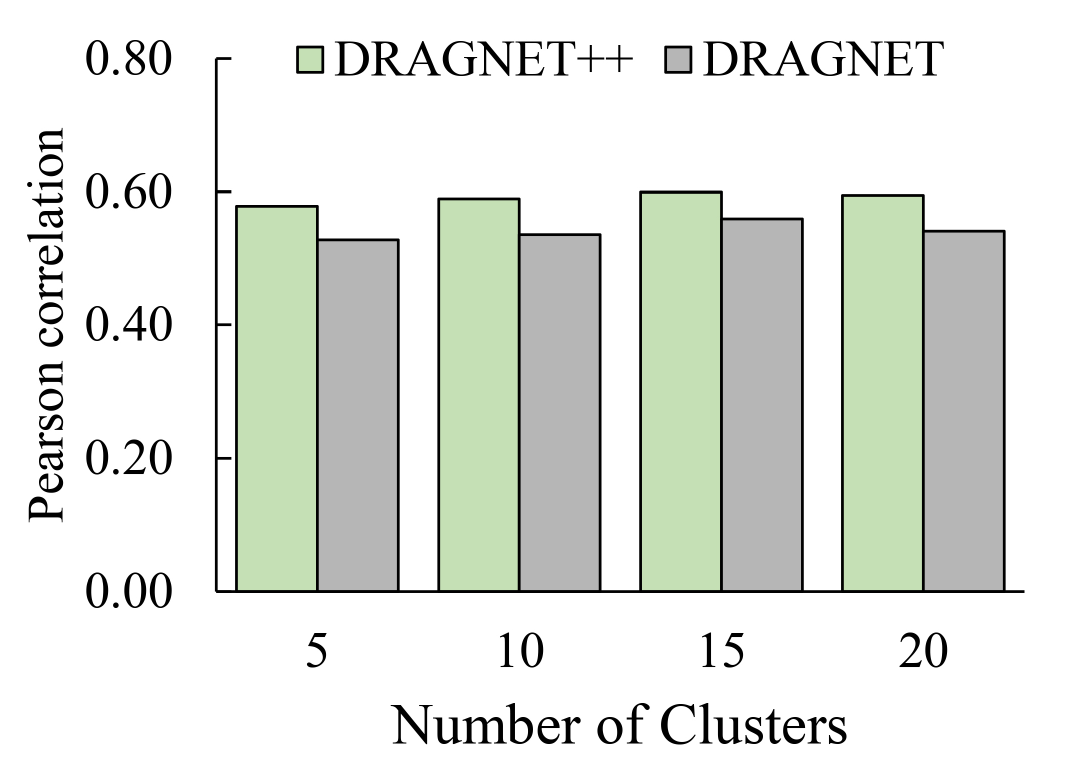}}
\qquad
\subfigure[Anti-Asian]{\includegraphics[width=0.28\textwidth]{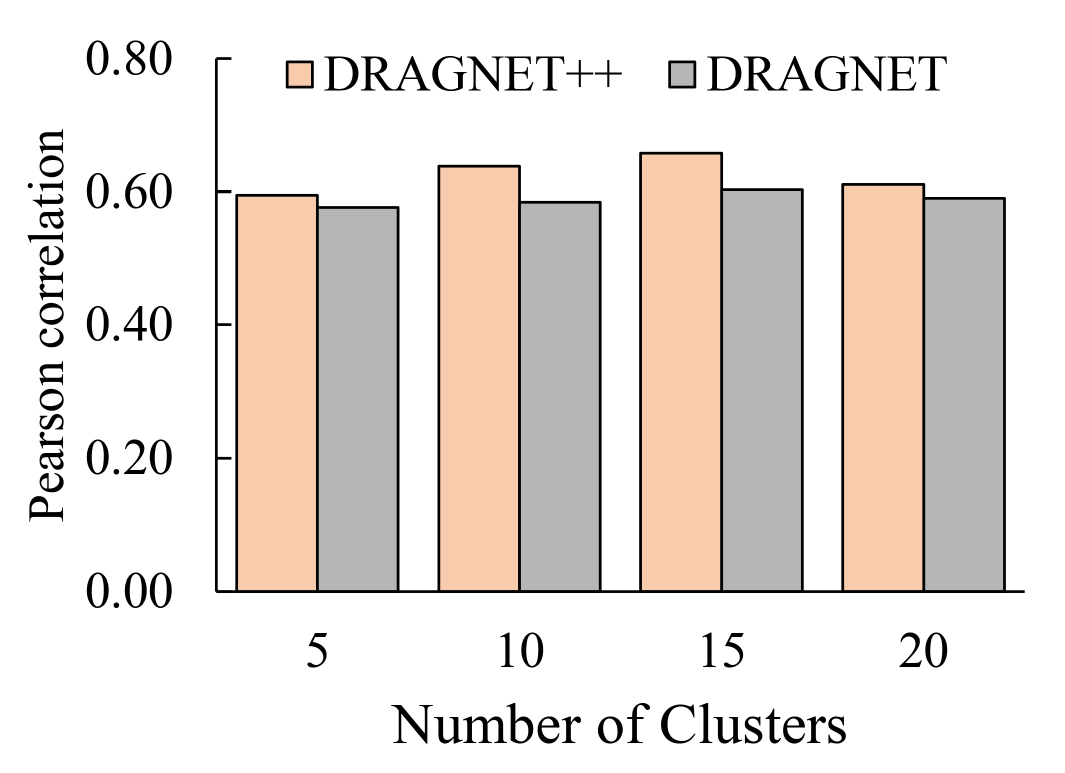}}
\caption{Performance comparison shown by changing the hyperparameter of number of clusters in fuzzy clustering algorithm.}
\label{fig:cluster_number}
\end{figure}

One of the most important hyperparameters of {\model} in the fuzzy clustering step is the number of clusters $j$.
As a result, we choose a number of clusters in the range of $\{5,10,15,20\}$ to assess the impact of cluster size. The experiment outcomes are shown in Fig \ref{fig:cluster_number}. We notice that as the number of clusters grows, the {\model}'s performance improves, and it is consistently better than {DRAGNET}. It is difficult to discern between different types of conversation threads when we use a small number of clusters. However, the performance begins to deteriorate beyond a certain number of clusters. One probable explanation is that grouping conversation threads into too many clusters produces noise, making it harder for the prior model to detect the correct cluster labels with minimal historical data. 


\subsubsection{The impact of different weights in hate intensity score}
\begin{figure}[!t]
\subfigure[Anti-Racism]{\includegraphics[width=0.28\textwidth]{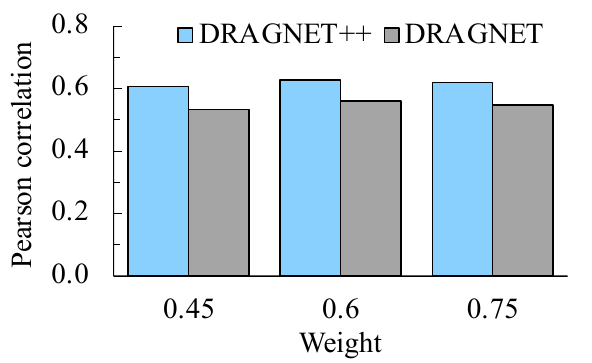}}
\qquad
\subfigure[Anti-Social]{\includegraphics[width=0.28\textwidth]{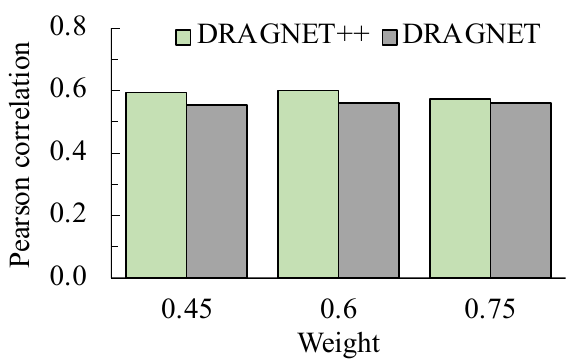}}
\qquad
\subfigure[Anti-Asian]{\includegraphics[width=0.28\textwidth]{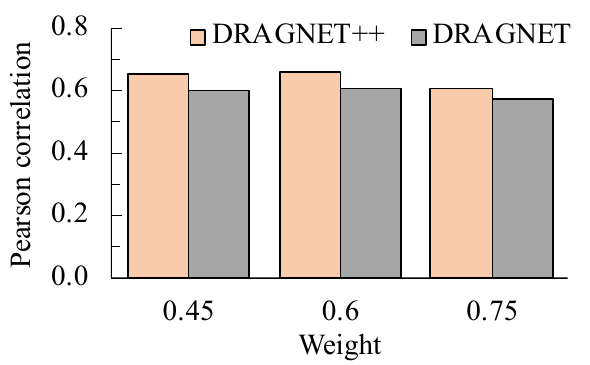}}
\caption{Performance comparison depicted by varying the weight (w) which optimizes the trade-off between hate detection/lexicon.}
\label{fig:hate_weight}
 \vspace{+5mm}
\end{figure}

Our hate intensity score leverages $w$ to simplify the trade-off between the hate detection model and the hate lexicon component in Section \ref{sec:preliminaries}. Our suggested {\model} model definitely outperforms {DRAGNET} for all three values of $w$ (i.e., 0.45, 0.6, 0.75) on all three datasets, as demonstrated in Fig \ref{fig:hate_weight}.

\subsubsection{The impact of hate speech detection models}
\begin{figure}[h]
\subfigure[Anti-Racism]{\includegraphics[width=0.28\textwidth]{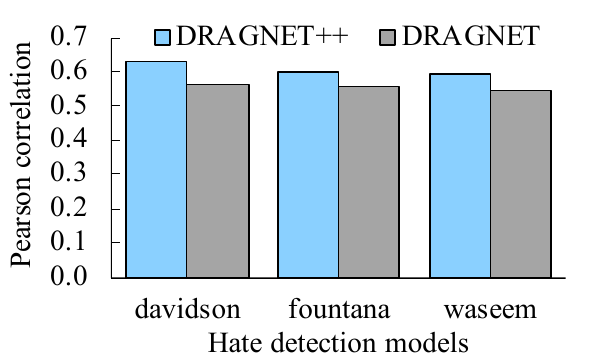}}
\qquad
\subfigure[Anti-Social]{\includegraphics[width=0.28\textwidth]{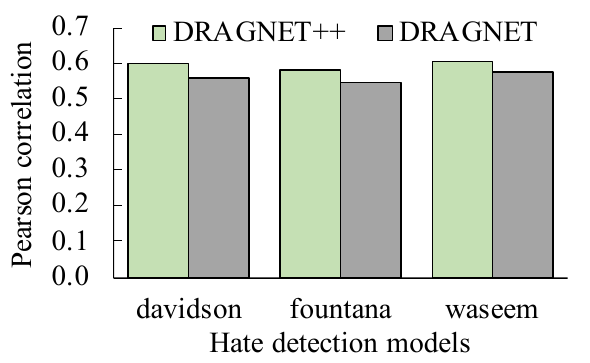}}
\qquad
\subfigure[Anti-Asian]{\includegraphics[width=0.28\textwidth]{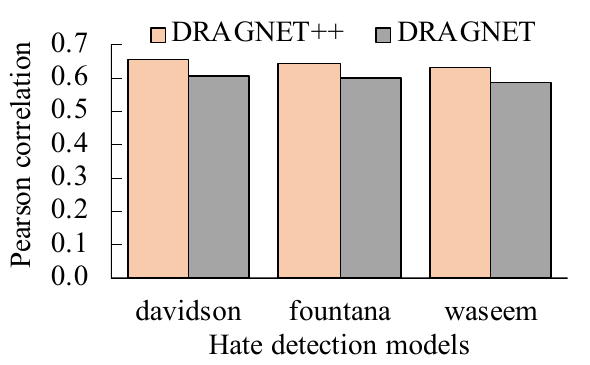}}
\caption{The impact of different hate detection algorithms on {\model} and {DRAGNET}}
\label{fig:detection_model}
\end{figure}

\red{During the data preprocessing phase, we utilized a hate speech detection model to compute hate intensity scores for replies. To examine the effect of different hate speech detection models, we evaluated three models proposed by \citet{davidson2017automated}] (model used in {\model}), \citet{founta2019unified}, and Waseem and Hovy \citet{waseem2016hateful}. As illustrated in Figure \ref{fig:detection_model}, the selected hate speech detection models had a minor impact on the ultimate performance. Nonetheless, our proposed model exhibited consistent superiority over DRAGNET across all three hate detection models.}

\section{Conclusion}
\label{sec:conclusion}
In this paper, we build on our previous work using two new datasets: \emph{Anti-Social} and \emph{Anti-Asian}. The new datasets gave enough bandwidth to continue experimenting with the old model {DRAGNET} to forecast hate intensity profiles. By adding the structural graph information critical for any Twitter reply chain, we suggested a new model, \model. GNN-augmented \model\ outperformed all baselines, including {DRAGNET}, by a large margin. We then used several ablations to compare {DRAGNET} and \model\ to figure out why our model outperformed the existing architecture.

In the future, we would like to build a meaningful association of hate intensity profiles at the branch level of the reply chain's tree structure using sophisticated structural graph information. We would also like to look into how user data can be used to predict hate intensity scores.
Human-annotated hate intensity datasets can be included. Our ultimate goal, sooner or later, would be to stop the spread of hate speech as soon as possible after the first tweet appears on Twitter.


\bibliographystyle{elsarticle-num-names} 
\bibliography{ref}





\end{document}